\documentclass[a4paper,prd,preprint,amsmath,amssymb,showpacs,nofootinbib]{revtex4}
\usepackage{graphicx}
\usepackage{color}
\definecolor{darkgreen}{rgb}{ .0, .8, .3}
\begin{document}

\title{New physics in the weak interaction of $\bar B\to D^{(*)}\tau\bar\nu$}


\author{Minoru TANAKA}
\email{tanaka@phys.sci.osaka-u.ac.jp}
\author{Ryoutaro WATANABE}
\email{ryoutaro@het.phys.sci.osaka-u.ac.jp}
\affiliation{Department of Physics, Graduate School of Science, Osaka University, Toyonaka, Osaka 560-0043, Japan}

\begin{abstract}
Recent experimental results on exclusive semi-tauonic $B$ meson decays, $\bar B\to D^{(*)}\tau\bar\nu$, 
showing sizable deviations from the standard model prediction, suggest a new physics in which the structure of the relevant weak charged interaction may differ from that of the standard model. 
We study the exclusive semi-tauonic $B$ decays in a model-independent manner using the most general set of four-Fermi interactions in order to clarify possible structures of the charged current in new physics. 
It turns out that correlations among observables including tau and $D^*$ polarizations and $q^2$ distributions are useful to distinguish possible new physics operators.
Further, we investigate some interesting models to exhibit the advantage of our model-independent analysis. 
As a result, we find that two Higgs doublet models without tree-level flavor changing neutral currents (FCNC) 
and the minimal supersymmetric standard model with $R$-parity violation are unlikely to explain the present experimental data, 
while two Higgs doublet models with FCNC and a leptoquark model are consistent with the data.
\end{abstract}


\preprint{OU-HET-772-2012}

\maketitle

\section{Introduction}
The $V-A$ structure of the weak interaction was established in the study of 
nuclear $\beta$ decays for the quarks and lepton in the first 
generation \cite{MRR}. 
Since then, several charged current interactions in the second and third 
generations have been investigated and turned out to be described 
by similar $V-A$ interactions with a reasonable level of accuracy.

Among the charged current processes observed so far, 
pure- and semi- tauonic $B$ meson decays, 
$\bar B\to\tau\bar\nu$ \cite{BABARPT,BELLEPT} and 
$\bar B\to D^{(*)}\tau\bar\nu$ \cite{BABARST,Babar2012,BELLEST1,BELLEST2,BELLEST3} are 
thought-provoking in the sense that they contain both the quark ($b$) 
and the lepton ($\tau$) in the third generation. 
Their large masses suggest a close connection with 
the electroweak symmetry breaking (EWSB), which is still not fully understood 
even after the discovery of a di-gamma resonance \cite{ATLAS,CMS}.

The EWSB sector in a candidate of new physics 
beyond the standard model (SM) often has a different structure
from that of the SM. A typical example is the two Higgs doublet model (2HDM)
of type II \cite{HHG}, which has the Higgs sector identical to that of 
the minimal supersymmetric standard model (MSSM) \cite{Nilles} 
at the tree level.
A pair of charged Higgs bosons appears in this class of models, and 
its couplings to fermions are proportional to the involved fermion masses 
and further enhanced if the ratio of vacuum expectation values, $\tan\beta$, 
is large. Several authors have studied effects of these enhanced 
couplings, which modify the $V-A$ structure, in the pure- 
and semi- tauonic $B$ decays\cite{GH,HOU93,TANAKA,KS,MT,MMT,IKO,CG2005,CG2006,NTW,TRINE,KMLAT,OUR,MILClattice}.

From the experimental point of view, these $B$ decay processes are 
rather difficult to identify because of two or more missing neutrinos
in the final states. With the help of large statistics and low backgrounds, 
however, they are good targets of $e^+e^-$ $B$ factory experiments.
The observed branching fraction of $\bar B\to\tau\bar\nu$ is 
$O(10^{-4})$ and that of $\bar B\to D^{(*)}\tau\bar\nu$ is $O(10^{-2})$. 
Comparing the pure- and semi- tauonic $B$ decays, the latter provides
a wide variety of observables besides the larger branching fraction, such as
decay distributions \cite{KS,CG2005,CG2006,NTW,TRINE,KMLAT,FKN,Sakaki,DDG, BKT,CJLP} and polarizations \cite{TANAKA,OUR,FKN,DDG}. 
Hence, the semi-tauonic processes allow us to investigate 
the structure of the relevant charged current interaction, 
and we concentrate on $\bar B\to D^{(*)}\tau\bar\nu$ in this work.

Another advantage of $\bar B\to D^{(*)}\tau\bar\nu$ is that
theoretical and experimental uncertainties tend to be reduced by
taking the ratio of branching fraction to the semileptonic decays,
$\bar B\to D^{(*)}\ell\bar\nu$, where $\ell$ denotes $e$ or $\mu$.
These decay modes including lighter charged leptons are supposed to be
less sensitive to the EWSB sector and assumed to be described by 
the SM in the present work. 

The ratios of the branching fractions are defined by
\begin{equation}
R(D^{(*)})\equiv\frac{\mathcal{B}(\bar B\to D^{(*)}\tau^-\bar\nu_\tau)}
                     {\mathcal{B}(\bar B\to D^{(*)}\ell^-\bar\nu_\ell)},
\end{equation}
and their experimental values are summarized as
\begin{equation}\label{Eq:combined}
R(D)=0.42\pm 0.06,\quad R(D^*)=0.34\pm 0.03, 
\end{equation}
where we combine the results of BaBar \cite{Babar2012} and Belle \cite{BELLEST1,BELLEST2,BELLEST3}  
assuming the gaussian distribution. These measurements seem to be
larger than the SM predictions,
\begin{equation}
R(D)=0.305\pm 0.012,\quad R(D^*)=0.252\pm 0.004,
\end{equation}
and the SM is disfavored at $3.5\sigma$.

Interestingly, as is shown in Ref.~\cite{Babar2012}, one of $R(D)$ and 
$R(D^*$) is made consistent with the corresponding experimental value 
in the 2HDM of type II by adjusting $\tan\beta/m_{H^\pm}$, 
but it is unlikely to make both of them agree with the experimental results.
Thus, the 2HDM of type II is also disfavored.

In the present work, we first provide a model-independent framework for 
the semi-tauonic $B$ decays in order to clarify the type of new physics 
that is required to explain the current experimental results.
Then, we apply the framework to analyzing several models. 
The rest of this paper is organized as follows: 
In Sec.~\ref{ELandHA}, we present the most general effective Lagrangian
of $b\to c\tau\bar\nu$ and the resultant helicity amplitudes of $\bar B\to D^{(*)} \tau\bar\nu$. 
Experimental constraints and theoretical predictions are given 
in Sec.~\ref{PredandConst}. We discuss how to distinguish new physics
contributions that appear in the effective Lagrangian in Sec.~\ref{Distinguish}.
We analyze 2HDMs, MSSM with $R$-parity violation and a leptoquark model in Sec.~\ref{Model}. Section \ref{Conclusion}
is devoted to our conclusions.

\section{Effective Lagrangian and Helicity Amplitudes}
\label{ELandHA}

\subsection{Effective Lagrangian for $b\to c\tau\bar\nu$}
The semi-tauonic bottom quark decay, $b\to c\tau\bar\nu$, is described by the four-Fermi interaction of charged left-handed currents in the SM. 
Other four-Fermi operators might be induced in the presence of new physics. 

The effective Lagrangian that contains all conceivable four-Fermi operators is written as
\begin{equation}
\label{Eq:EL}
 -\mathcal{L}_\text{eff} = 2\sqrt{2}G_F V_{cb}\sum_{l=e,\mu,\tau}\left[ (\delta_{l\tau} +C_{V_1}^l)\mathcal{O}_{V_1}^l +C_{V_2}^l\mathcal{O}_{V_2}^l 
  +C_{S_1}^l\mathcal{O}_{S_1}^l +C_{S_2}^l\mathcal{O}_{S_2}^l +C_T^l\mathcal{O}_T^l\right],
\end{equation}
where the four-Fermi operators are defined by
\begin{eqnarray}
 \mathcal{O}_{V_1}^l&=&\bar{c}_L\gamma^\mu b_L\, \bar{\tau}_L\gamma_\mu \nu_{Ll}\,, \label{Eq:OV1} \\
 \mathcal{O}_{V_2}^l&=&\bar{c}_R\gamma^\mu b_R\, \bar{\tau}_L\gamma_\mu \nu_{Ll}\,,\\
 \mathcal{O}_{S_1}^l&=&\bar{c}_L b_R\,\bar{\tau}_R \nu_{Ll}\,,\\ 
 \mathcal{O}_{S_2}^l&=&\bar{c}_R b_L\,\bar{\tau}_R \nu_{Ll}\,,\\ 
 \mathcal{O}_T^l&=&\bar{c}_R\sigma^{\mu\nu}b_L\, \bar{\tau}_R\sigma_{\mu\nu}\nu_{Ll}\,, \label{Eq:OT}
\end{eqnarray}
and $C_X^l\,(X=V_{1,2},S_{1,2},T)$ denotes the Wilson coefficient of $\mathcal{O}_X^l$.
Here we assume that the light neutrinos are left-handed.%
\footnote{Possibilities to introduce a light right-handed neutrino are discussed in Refs.~\cite{HeValencia,FKNZ}.}
The neutrino flavor is specified by $l$, and we take all cases of $l=e$, $\mu$ and $\tau$ into account in the contributions of new physics. 
Since the neutrino flavor is not observed in the experiments of bottom decays, 
the neutrino mixing does not affect the following argument provided that the Pontecorvo-Maki-Nakagawa-Sakata matrix is unitary. 
The SM contribution is expressed by the term of $\delta_{l\tau}$ in Eq.~(\ref{Eq:EL}). 
We note that the tensor operator with the opposite set of quark chiralities identically vanishes:
$\bar{c}_L\sigma^{\mu\nu}b_R\,\bar{\tau}_R\sigma_{\mu\nu}\nu_{Ll}=0$.

\subsection{Helicity amplitudes}
The helicity amplitude of $\bar B\to M\tau\bar\nu\,(M=D,D^*)$ is written as 
\begin{equation}
 \mathcal{M}^{\lambda_\tau,\lambda_M}_l
 =\delta_{l\tau}\,\mathcal{M}^{\lambda_\tau,\lambda_M}_\text{SM} +\mathcal{M}^{\lambda_\tau,\lambda_M}_{V_1,l} +\mathcal{M}^{\lambda_\tau,\lambda_M}_{V_2,l}
 +\mathcal{M}^{\lambda_\tau,\lambda_M}_{S_1,l} +\mathcal{M}^{\lambda_\tau,\lambda_M}_{S_2,l} +\mathcal{M}^{\lambda_\tau,\lambda_M}_{T,l} \,,
\end{equation}
where $\lambda_\tau$ is the helicity of the tau lepton, $\lambda_M=s$ indicates the amplitude of $\bar B\to D\tau\bar\nu$, and $\lambda_M=\pm1,0$ denotes the $D^*$ helicity defined in the $B$ rest frame . 
The amplitude $\mathcal{M}^{\lambda_\tau,\lambda_M}_\text{SM}$ represents the SM contribution 
and other terms in the right-hand side stand for new physics contributions corresponding to the operators in Eqs.~(\ref{Eq:OV1})--(\ref{Eq:OT}).
The SM amplitude is given by \cite{HMW89,HMW90}
\begin{equation}
 \mathcal{M}^{\lambda_\tau,\lambda_M}_\text{SM}
 =\frac{G_F}{\sqrt{2}}V_{cb} \sum_{\lambda}\eta_{\lambda}H_{V_1,\lambda}^{\lambda_M}L_{\lambda,\tau}^{\lambda_\tau}\,,
\end{equation}
and the amplitudes that represent new physics contributions take the following forms:
\begin{eqnarray}
 \mathcal{M}^{\lambda_\tau,\lambda_M}_{V_i,l}
 &=& \frac{G_F}{\sqrt{2}}V_{cb}C_{V_i}^l \sum_{\lambda}\eta_{\lambda}H_{V_i,\lambda}^{\lambda_M}L_{\lambda,l}^{\lambda_\tau} \hspace{1em} (i=1,2)\,,\\
 \mathcal{M}^{\lambda_\tau,\lambda_M}_{S_i,l} 
 &=& -\frac{G_F}{\sqrt{2}}V_{cb}C_{S_i}^l H_{S_i}^{\lambda_M}L_l^{\lambda_\tau} \hspace{1em} (i=1,2)\,,\\
  \mathcal{M}^{\lambda_\tau,\lambda_M}_{T,l}
 &=& -\frac{G_F}{\sqrt{2}}V_{cb}C_T^l \sum_{\lambda,\lambda'}\eta_\lambda\eta_{\lambda'} H_{\lambda\lambda'}^{\lambda_M}L_{\lambda\lambda',l}^{\lambda_\tau}\,,
\end{eqnarray}
where $H$'s and $L$'s are the hadronic and leptonic amplitudes respectively as defined below; 
$\lambda,\lambda'=\pm,0,s$ denote the virtual vector boson helicity. The metric factor $\eta_{\lambda}$ is given by $\eta_{\pm,0}=1$ and $\eta_s=-1$. 
We treat the contraction of the Lorentz indices in $\mathcal{O}_T^l$ as if two heavy vector bosons are exchanged. 

The leptonic amplitudes, $L_{\lambda,l}^{\lambda_\tau}$, $L_l^{\lambda_\tau}$ and $L_{\lambda\lambda',l}^{\lambda_\tau}$ are defined by
\begin{eqnarray}
\label{Eq:HALV}
 L_{\lambda,l}^{\lambda_\tau}(q^2,\cos\theta_\tau) 
 &=& \epsilon_\mu(\lambda) \langle\tau(p_\tau,\lambda_\tau)\bar\nu_l(p_\nu)| \bar\tau\gamma^\mu(1-\gamma_5)\nu_l|0\rangle\,,\\
\label{Eq:HALS}
 L_l^{\lambda_\tau}(q^2,\cos\theta_\tau) 
 &=& \langle\tau(p_\tau,\lambda_\tau)\bar\nu_l(p_\nu)| \bar\tau(1-\gamma_5)\nu_l|0\rangle\,,\\
\label{Eq:HALT}
 L_{\lambda\lambda',l}^{\lambda_\tau}(q^2,\cos\theta_\tau) 
 &=& -i\epsilon_\mu(\lambda)\epsilon_\nu(\lambda') \langle\tau(p_\tau,\lambda_\tau)\bar\nu_l(p_\nu)| \bar\tau\sigma^{\mu\nu}(1-\gamma_5)\nu_l|0\rangle\,,
\end{eqnarray}
where $\epsilon_\mu(\lambda)$ represents the polarization vector of the virtual vector boson, 
$q^\mu=p_B^\mu-p_M^\mu(=p_\tau^\mu+p_\nu^\mu)$ is the momentum transfer, 
and $\theta_\tau$ denotes the angle between the momentum of the tau lepton and that of the $M$ meson 
in the rest frame of the leptonic system, to which we refer as the $q$ rest frame \cite{OUR}.
The $\tau$ helicity $\lambda_\tau$ is also defined in the $q$ rest frame. The explicit formulae of the leptonic amplitudes are relegated to Appendix \ref{Ap:LA}.

The hadronic amplitudes, $H_{V_i,\lambda}^{\lambda_M}$, $H_{S_i}^{\lambda_M}$ and $H_{\lambda \lambda'}^{\lambda_M}$ are defined by
\begin{eqnarray}
\label{Eq:HAV1}
 H_{V_1,\lambda}^{\lambda_M}(q^2)
 &=&\epsilon^*_\mu(\lambda) \langle M(p_M,\epsilon \left(\lambda_M)\right)| \bar c \gamma^\mu(1-\gamma^5) b| \bar B (p_B)\rangle\,,\\
\label{Eq:HAV2}
 H_{V_2,\lambda}^{\lambda_M}(q^2)
 &=&\epsilon^*_\mu(\lambda) \langle M(p_M,\epsilon \left(\lambda_M)\right)| \bar c \gamma^\mu(1+\gamma^5) b| \bar B (p_B)\rangle\,,\\
\label{Eq:HAS1}
 H_{S_1}^{\lambda_M}(q^2)
 &=&\langle M(p_M,\epsilon \left(\lambda_M)\right)| \bar c (1+\gamma^5) b| \bar B (p_B)\rangle\,,\\
\label{Eq:HAS2}
 H_{S_2}^{\lambda_M}(q^2)
 &=&\langle M(p_M,\epsilon \left(\lambda_M)\right)| \bar c (1-\gamma^5) b| \bar B (p_B)\rangle\,,\\
\label{Eq:HAT}
 H_{\lambda\lambda'}^{\lambda_M}(q^2)
 &=&i \epsilon^*_\mu(\lambda)\epsilon^*_\nu(\lambda') \langle M(p_M,\epsilon \left(\lambda_M)\right)| \bar c \sigma^{\mu\nu}(1-\gamma^5) b| \bar B (p_B)\rangle,
\end{eqnarray}
where $\epsilon(\lambda_M)$ denotes the polarization vector of $D^*$ for $\lambda_M=\pm1,0$.   
The relations $H_{V_1,\lambda}^s=H_{V_2,\lambda}^s$ and  $H_{S_1}^s=H_{S_2}^s$ hold because of the parity conservation in the strong interaction.  
Similarly, we find $H_{S_1}^{\lambda_M}=-H_{S_2}^{\lambda_M}$ for $\lambda_M=\pm1,0$. 

The hadronic amplitudes of the vector type operators for $\bar B \to D \tau \bar\nu$ defined in Eqs.~(\ref{Eq:HAV1}) and (\ref{Eq:HAV2}) are represented as
\begin{eqnarray}
 \label{Eq:HAVresult1} H_{V_1,\pm}^s(q^2) &=& H_{V_2,\pm}^s(q^2) =0 \,,\\
 H_{V_1,0}^s(q^2) &=& H_{V_2,0}^s(q^2) =m_B \sqrt{r} (1+r) \frac{\sqrt{w^2-1}}{\sqrt{\hat q^2(w)}} \,V_1(w) \,,\\
 H_{V_1,s}^s(q^2) &=& H_{V_2,s}^s(q^2) =m_B \sqrt{r} (1-r)  \frac{w+1}{\sqrt{\hat q^2(w)}}\,S_1(w) \,, 
\end{eqnarray}
and those for $\bar B \to D^* \tau \bar\nu$ are 
\begin{eqnarray}
 H_{V_1,\pm}^\pm(q^2) &=& -H_{V_2,\mp}^\mp(q^2) =m_B \sqrt{r} A_1(w) \left [ w+1 \mp \sqrt{w^2-1} R_1(w) \right ] \,,\\
 H_{V_1,0}^0(q^2) &=& -H_{V_2,0}^0(q^2) =m_B \sqrt{r} \frac{w+1}{\sqrt{\hat q^2(w)}} A_1(w) \left [ -w+r +(w-1) R_2(w) \right ] \,,\\
 H_{V_1,s}^0(q^2) &=& -H_{V_2,s}^0(q^2) \notag \\
 &=&\frac{m_B}{2\sqrt{r}} \frac{\sqrt{w^2-1}}{\sqrt{\hat q^2(w)}} A_1(w) \left [ -2r(w+1) +(1-r^2) R_2(w) -\hat q^2(w) R_3(w) \right ] \,,  \label{Eq:HAVresult2} \\
 \text{others} &=& 0,
\end{eqnarray}
where $\hat q^2(w)=1+r^2-2rw$, $r=m_M/m_B$, and $w=p_B\cdot p_M/(m_Bm_M)$ is the velocity transfer in $\bar B\to M\tau\bar\nu$. 
The form factors, $V_1(w), S_1(w), A_1(w)$ and $R_i(w)$ are defined in Appendix \ref{Ap:HA}. 

The amplitudes of the scalar type operators are expressed in terms of vector form factors by applying the equations of motion of the quark fields: 
\begin{eqnarray}
 H_{S_1}^s(q^2) &=& H^s_{S_2}(q^2) = m_B\sqrt{r}(w+1)S_1(w)\,, \\
 H_{S_1}^\pm(q^2) &=& H_{S_2}^\pm(q^2) = 0\,, \\
 H_{S_1}^0(q^2) &=& -H_{S_2}^0(q^2) \notag \\
 &=& \frac{m_B \sqrt{w^2-1}}{2\sqrt{r}(1+r)} A_1(w) \left [-2r(w+1) +(1-r^2)R_2(w) -\hat q^2(w) R_3(w) \right ].
\end{eqnarray}
Similarly, for the tensor operator, we obtain  
\begin{eqnarray}
 H_{+-}^s(q^2) &=& H_{0s}^s(q^2) = m_B\sqrt{r}\frac{\sqrt{w^2-1}}{\hat q^2(w)} [-(1+r)^2V_1(w)+2r(w+1)S_1(w)]\,, \\
 H_{\pm0}^\pm(q^2) 
 &=& \pm H_{\pm s}^\pm(q^2) \notag \\
 &=& \frac{ m_B \sqrt{r} }{ \sqrt {\hat q^2(w)} } A_1(w) \left [ \pm (1-r) (w+1) +(1+r) \sqrt{w^2-1} R_1(w) \right ] \,,\\
 H_{+-}^0(q^2) 
 &=& H_{0s}^0(q^2) \notag \\
 &=& \frac{m_B \sqrt{r}}{1+r} A_1(w) \left [ -(w+r)(w+1) +(w^2-1) R_3(w) \right ] \,, \\ 
 H_{\lambda\lambda'}^{\lambda_M}(q^2) &=& -H_{\lambda'\lambda}^{\lambda_M}(q^2) \,,
\end{eqnarray} 
and $\text{others} = 0$.  
A detailed derivation of these hadronic amplitudes is found in Appendix~\ref{Ap:HA}. 

\subsection{Form factors}
\label{Sec:FF}
The form factor $V_1(w)$ is extracted from the experimental data on $\bar B\to D\ell\bar\nu$, where $\ell=e,\mu$, provided that the decay process is described by the SM. 
Employing the following ansatz \cite{CLN}, 
\begin{equation}
 V_1(w)=V_1(1)\left[1-8\rho_1^2 z+(51.\rho_1^2-10.)z^2 -(252.\rho_1^2-84.)z^3 \right]\,,
\end{equation}
where $z=(\sqrt{w+1}-\sqrt{2})/(\sqrt{w+1}+\sqrt{2})$, the heavy flavor averaging group (HFAG) determines the slope parameter $\rho_1^2$ as $\rho_1^2=1.186\pm 0.055$ \cite{HFAG}.

The form factor $S_1(w)$ does not contribute to the semileptonic $B$ decay into a massless charged lepton. 
However it reduces to the Isgur-Wise function \cite{IW} in the heavy quark limit as well as $V_1(w)$. 
Accordingly we parametrize  it as 
\begin{equation}
 S_1(w)= \left[1+\Delta(w)\right]V_1(w)\,,
\end{equation}
where $\Delta(w)$ denotes the QCD and $1/m_Q$ corrections.
Following Refs.~\cite{MMT,CLN,NEUBERT94}, we estimate $\Delta(w)$ and give an approximate formula \cite{OUR}:
\begin{equation}\label{Eq:NLO}
 \Delta(w)=-0.019+0.041(w-1)-0.015(w-1)^2\,.
\end{equation}

As for $\bar B\to D^*\ell\bar\nu$, three form factors, $A_1(w)$ and $R_{1,2}(w)$, contribute. 
They are extracted from experimental data using the following parametrizations: 
\begin{eqnarray}
\label{Eq:A1}
 A_1(w) &=& A_1(1)\left [1-8\rho_{A_1}^2 z +(53\rho_{A_1}^2-15)z^2 -(231\rho_{A_1}^2-91)z^3\right ] \,,\\
\label{Eq:R1}
 R_1(w) &=& R_1(1) -0.12(w-1) +0.05(w-1)^2 \,,\\
\label{Eq:R2}
 R_2(w) &=& R_2(1) +0.11(w-1) -0.06(w-1)^2 \,.
\end{eqnarray}
Eq.~(\ref{Eq:A1}) is given in Ref.\,\cite{CLN} as well as $V_1(w)$. 
The $w$ dependence of $R_{1,2}$ is estimated by the heavy quark effective theory\cite{NEUBERT94}, while $R_1(1)$ and $R_2(1)$ are left as fitting parameters. 
The HFAG determines these parameters as follows \cite{HFAG}: $\rho_{A_1}^2=1.207\pm 0.026, R_1(1)=1.403\pm 0.033$ and $R_2(1)=0.854\pm 0.020$. 
The form factor $R_3(w)$ appears only in $\bar B\to D^*\tau\bar\nu$ and is estimated by the heavy quark effective theory as \cite{CLN}  
\begin{equation}
 R_3(w) = 1 + \Delta_3(w),\quad
 \Delta_3(w) =0.22 -0.052(w-1) +0.026(w-1)^2\,.
\end{equation}

\section{Experimental constraints and theoretical predictions}
\label{PredandConst}

\subsection{Observables and experimental constraints}
\label{Obs}
There are several measurable quantities affected by new physics operators in $\bar B\to D^{(*)}\tau\bar\nu$.
We consider  the decay rate and the tau longitudinal polarization in $\bar B\to D\tau\bar\nu$ and define the following quantities: 
\begin{equation}
 R (D) =\frac{\Gamma^+(D) +\Gamma^-(D)}{\Gamma(\bar B \to D \ell\bar\nu)} \,, \ \ 
 P_\tau (D) = \frac{\Gamma^+(D) -\Gamma^-(D)}{\Gamma^+(D) +\Gamma^-(D)}\,,
\end{equation}
where $\Gamma^\pm(D)$ represents the decay rate with $\lambda_\tau =\pm1/2$ as defined in Eq.~(\ref{Eq:tauhelicity}) and $\Gamma(\bar B \to D \ell\bar\nu)$ is the total decay rate of $\bar B \to D \ell\bar\nu$.
The uncertainties that come from $V_{cb}$ and the normalization factor $V_1(1)$ vanish in the above quantities. 
For $\bar B\to D^*\tau\bar\nu$ we also define $R(D^*)$ and $P_\tau(D^*)$ in the same fashion. 
In addition we introduce the $D^*$ polarization as 
\begin{equation}
 P_{D^*} = \frac{\Gamma(D^*_L)}{\Gamma(D^*_L) +\Gamma(D^*_T)}\,,
\end{equation}
where $D^*_{L(T)}$ represents the longitudinally (transversely) polarized $D^*$ and $\Gamma(D^*_{L,T})$ are defined in Eq.~(\ref{Eq:Dstar}). 
The uncertainties in $R(D^*)$, $P_\tau(D^*)$ and $P_{D^*}$ due to $V_{cb}$ and $A_1(1)$ also disappear. 
The new physics operators are expected to affect these observables in various ways. 
Thus it is important to study them at the same time in order to distinguish the underlying new physics. 

Besides the above integrated quantities, $q^2$ distributions are potentially sensitive to new physics. 
As we will illustrate below, the $q^2$ distribution of $\bar B \to D \tau \bar \nu$ decay rate is helpful in discriminating between two scalar operators.  

In the following model-independent analysis, we assume that one of the new physics operators in Eq.~(\ref{Eq:EL}) is dominant except the SM contribution. 
This assumption allows us to determine the dominant Wilson coefficient from the experimental results of $R(D)$ and $R(D^*)$, and to predict other observables. 
A situation beyond this assumption is discussed in Sec.~\ref{Model}.

\begin{figure}[!ht]
\includegraphics[viewport=0 5 1466 394,width=35em]{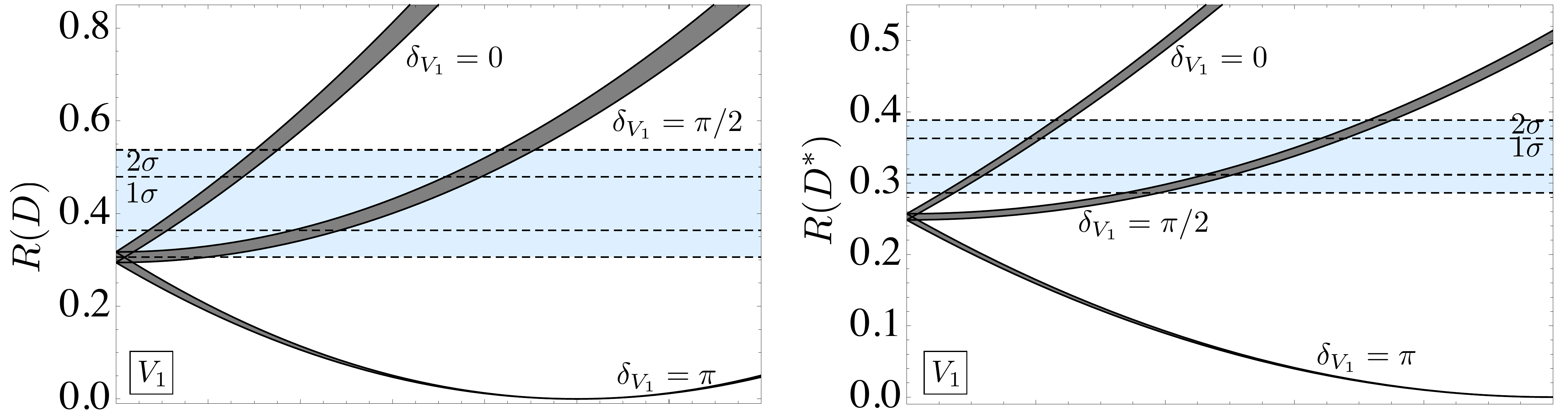} \\
\includegraphics[viewport=0 5 1466 394,width=35em]{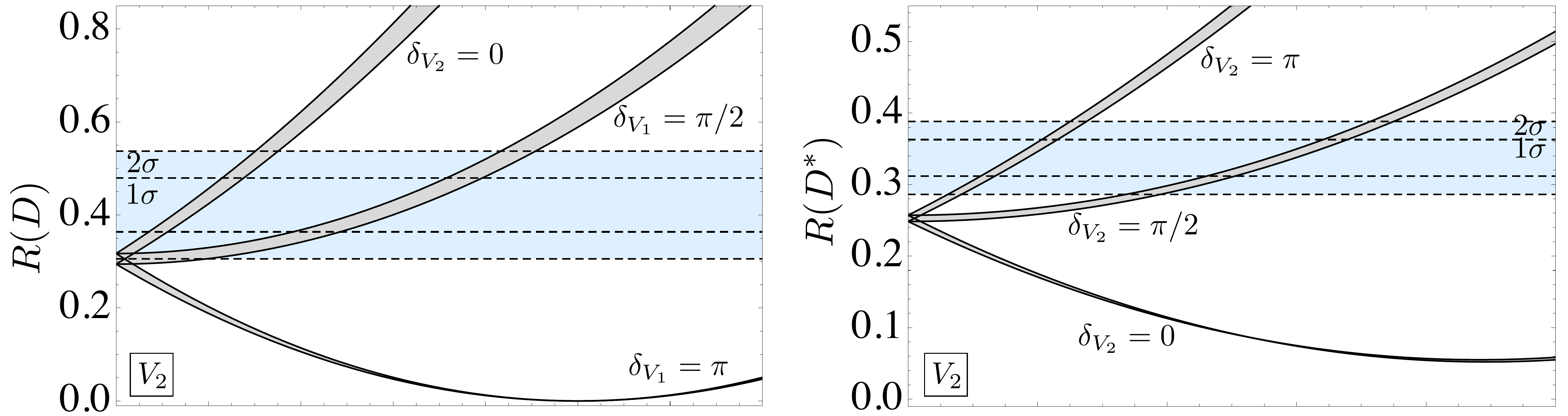} \\
\includegraphics[viewport=0 5 1466 392,width=35em]{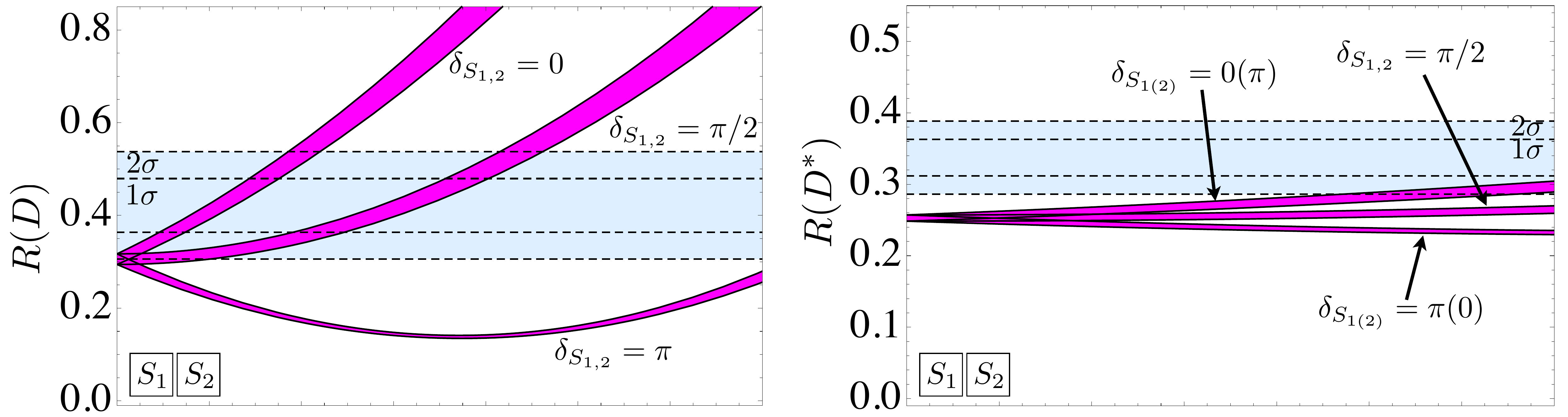} \\
\includegraphics[viewport=0 5 1466 475,width=35em]{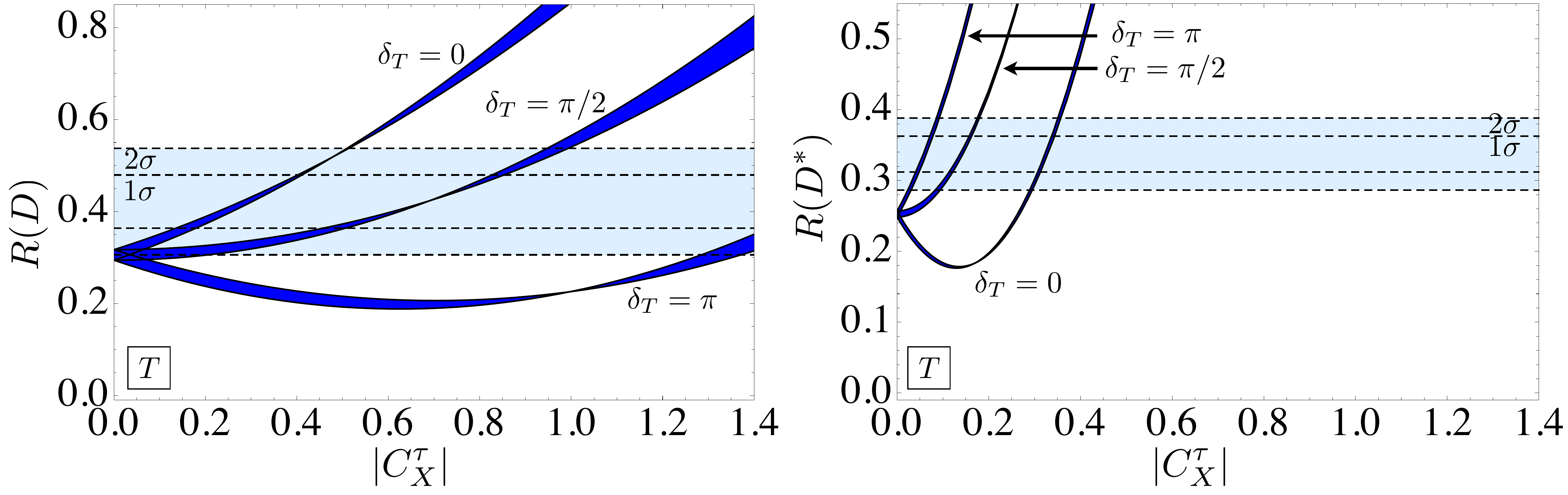} 
\caption{Predictions on the branching ratios as functions of the absolute value of Wilson coefficient $\left| C_X^\tau \right|$ for $X=V_{1,2},S_{1,2},T$.
The predictions of new physics effects for the operators $\mathcal O_X^{e,\mu}$ are given by the lines for $\delta_X=\pi/2$ in these graphs. 
The light blue horizontal bands represent the experimental values. }
\label{FIG:Rratios}
\end{figure}
\begin{figure}[!ht]
\includegraphics[viewport=0 0 1679 283,width=36em]{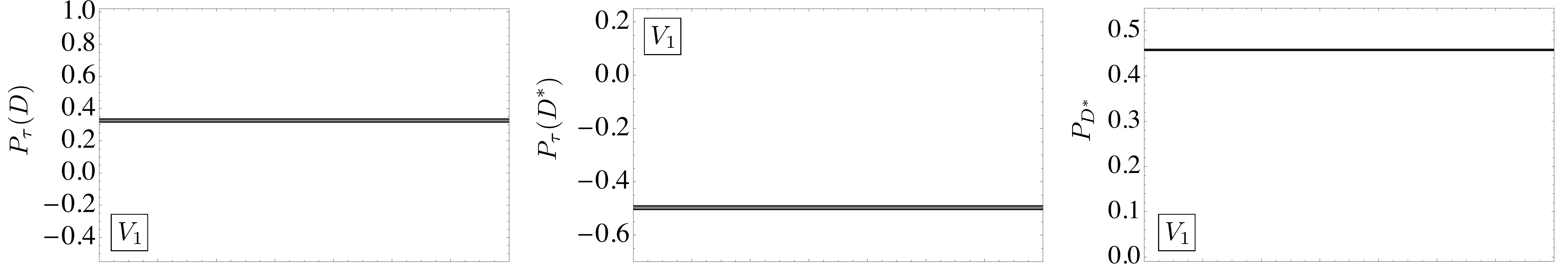} \\
\includegraphics[viewport=0 0 1679 283,width=36em]{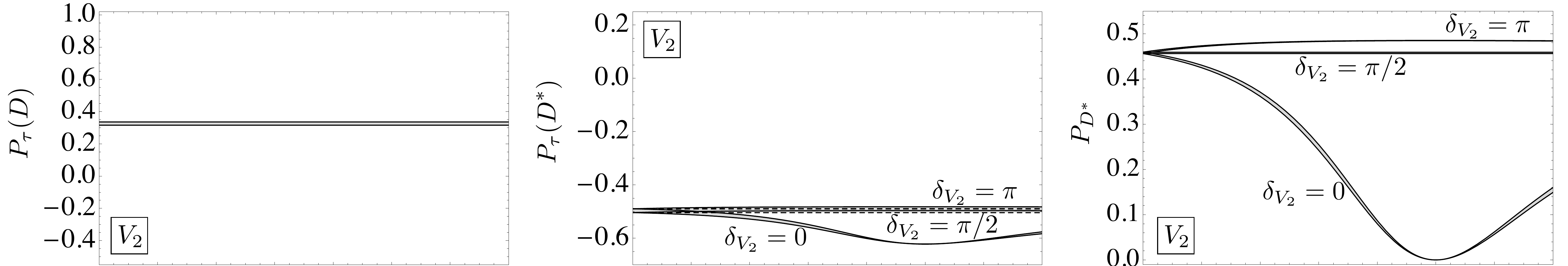} \\
\includegraphics[viewport=0 0 1679 283,width=36em]{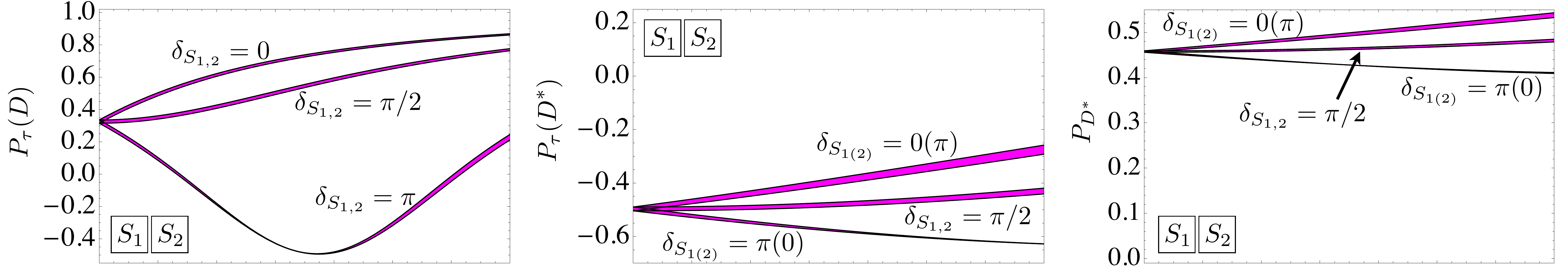} \\
\includegraphics[viewport=0 0 1679 355,width=36em]{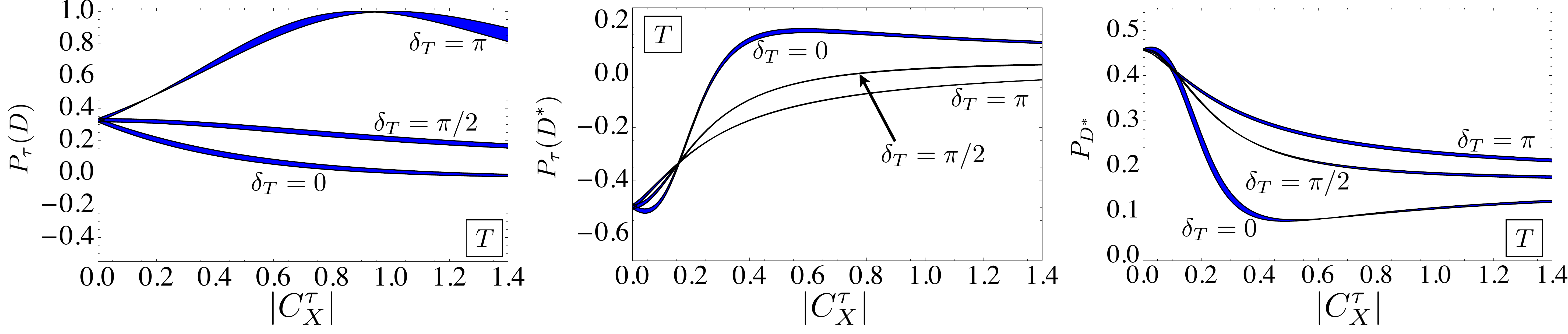} 
\caption{Predictions on $P_\tau(D),P_\tau(D^*)$ and $P_{D^*}$ as functions of $\left| C_X^\tau \right|$ for $X=V_{1,2},S_{1,2},T$.
The predictions of new physics effects for the operators $\mathcal O_X^{e,\mu}$ are given by the lines for $\delta_X=\pi/2$.}
\label{FIG:Pola}
\end{figure}

In Figs.~\ref{FIG:Rratios} and \ref{FIG:Pola}, we show new physics effects on $R(D^{(*)})$, $P_\tau(D^{(*)})$ and $P_{D^*}$. 
The horizontal axis is $|C^\tau_X|$ and three cases of $\delta_X=0, \pi /2$, and $\pi$ are shown for illustration, where $\delta_X$ is the complex phase of $C_X^\tau$.  
Effects of $\mathcal O^{e,\mu}_X$ are the same as those of $\mathcal O^\tau_X$ with $\delta_X=\pi /2$ because the new physics contributions do not interfere with the SM amplitudes in these cases. 
The width of each prediction indicates uncertainties due to the form factors. 
We include $\pm100$\% errors in the overall magnitudes of $\Delta(w)$ and $\Delta_3(w)$ as uncertainties in addition to the ranges of $\rho_1^2, \rho_{A_1}^2, R_1(1)$ and $R_2(1)$. 
The horizontal bands with dashed boundaries in Fig.~\ref{FIG:Rratios} represent the experimental values given in Eq.~(\ref{Eq:combined}). 
From these results, we find that the sensitivity to the magnitude of the Wilson coefficient varies depending on each operator. 
We note that the theoretical uncertainties are sufficiently smaller than the present experimental accuracy. 
Therefore, we use the central values of the theoretical predictions in the rest of this work unless otherwise stated. 

\begin{figure}[!ht]
\includegraphics[viewport=0 0 1402 434,width=36em]{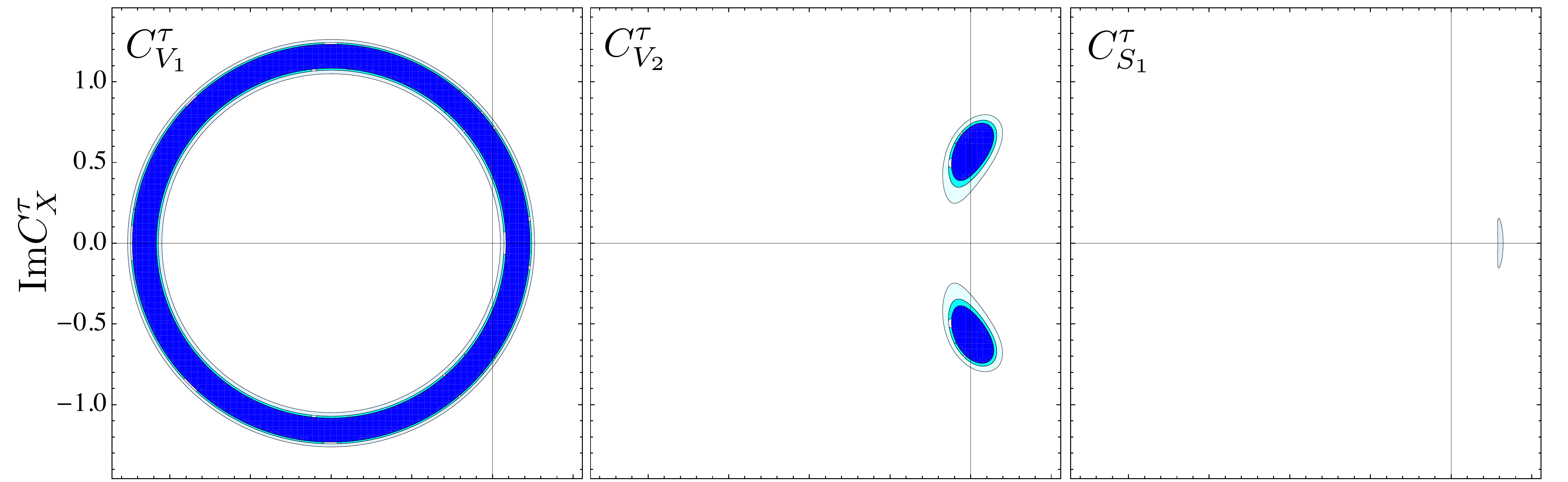} \\
\includegraphics[viewport=0 5 1402 501,width=36em]{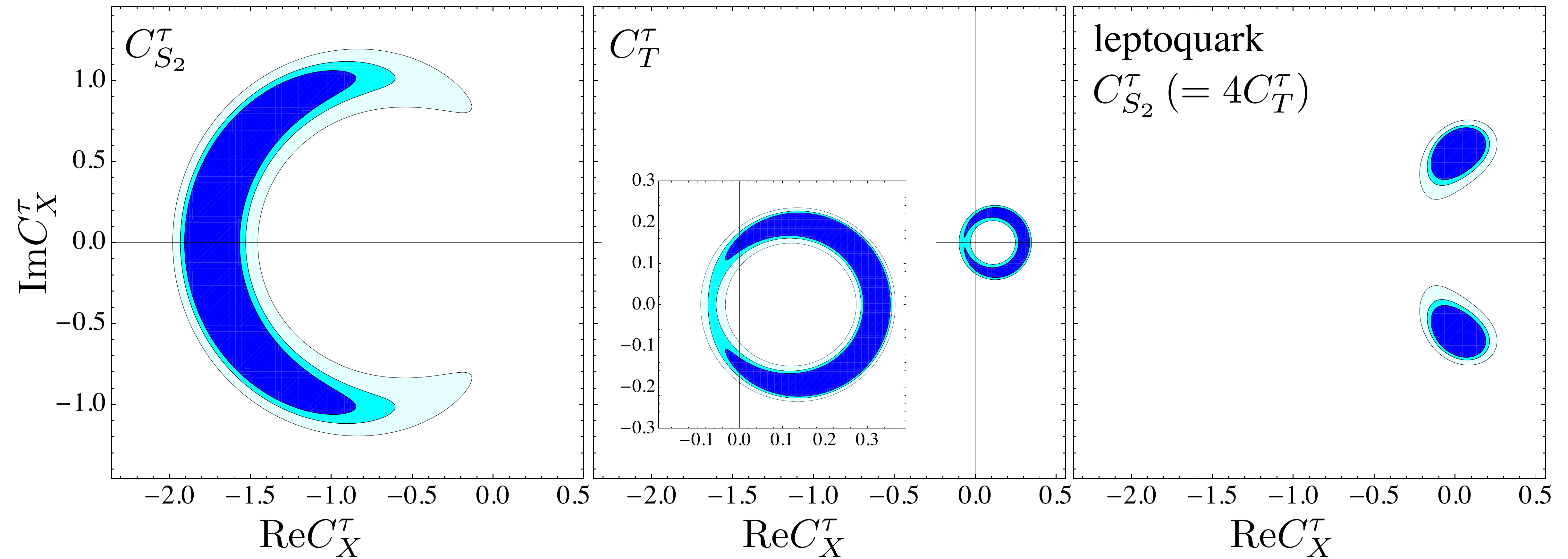} 
\caption{Constraints on the Wilson coefficient $C_X^\tau$. 
The allowed regions within $90\%$(blue), $95\%$(cyan) and $99\%$(light blue) C.L. are shown. 
Allowed regions for $\left| C_X^{e,\mu} \right|$ are obtained by those for pure imaginary $C_X^\tau$. 
The leptoquark panel is mentioned in Sec.~\ref{LQ}.}
\label{FIG:Contour}
\end{figure}
In Fig.~\ref{FIG:Contour}, the allowed region of the complex Wilson coefficient $C_X^\tau$ is shown for each operator $\mathcal O_X^\tau$.
The vector operators $\mathcal O_{V_{1,2}}^\tau$ describe the current experimental results \cite{FKNZ}. 
The operator $\mathcal O_{S_1}^\tau$ is unlikely while $\mathcal O_{S_2}^\tau$ is favored as is already pointed out in Refs.~\cite{FKNZ,CGK}. 
In addition, we find that the tensor operator $\mathcal O_T^\tau$ reasonably explains the current data. 
We can read the allowed regions of $\left|C_X^{e,\mu}\right|$ from that of the pure imaginary $C_X^\tau$, 
since neither $C_X^{e,\mu}$ nor the pure imaginary $C_X^\tau$ interferes with the SM contribution as is mentioned above. 
Thus, we find no allowed region of the Wilson coefficient within $99\%$ confidence level (C.L.) for $\mathcal O_{S_1}^{e,\mu}$ nor $\mathcal O_{S_2}^{e,\mu}$. 
The operators $\mathcal O_{V_1}^{e,\mu}, \mathcal O_{V_2}^{e,\mu}$ and $\mathcal O_T^{e,\mu}$ are able to explain the present data.

\subsection{Correlation between $R(D)$ and $R(D^*)$}
\begin{figure}
\includegraphics[viewport=0 0 1104 513,width=36em]{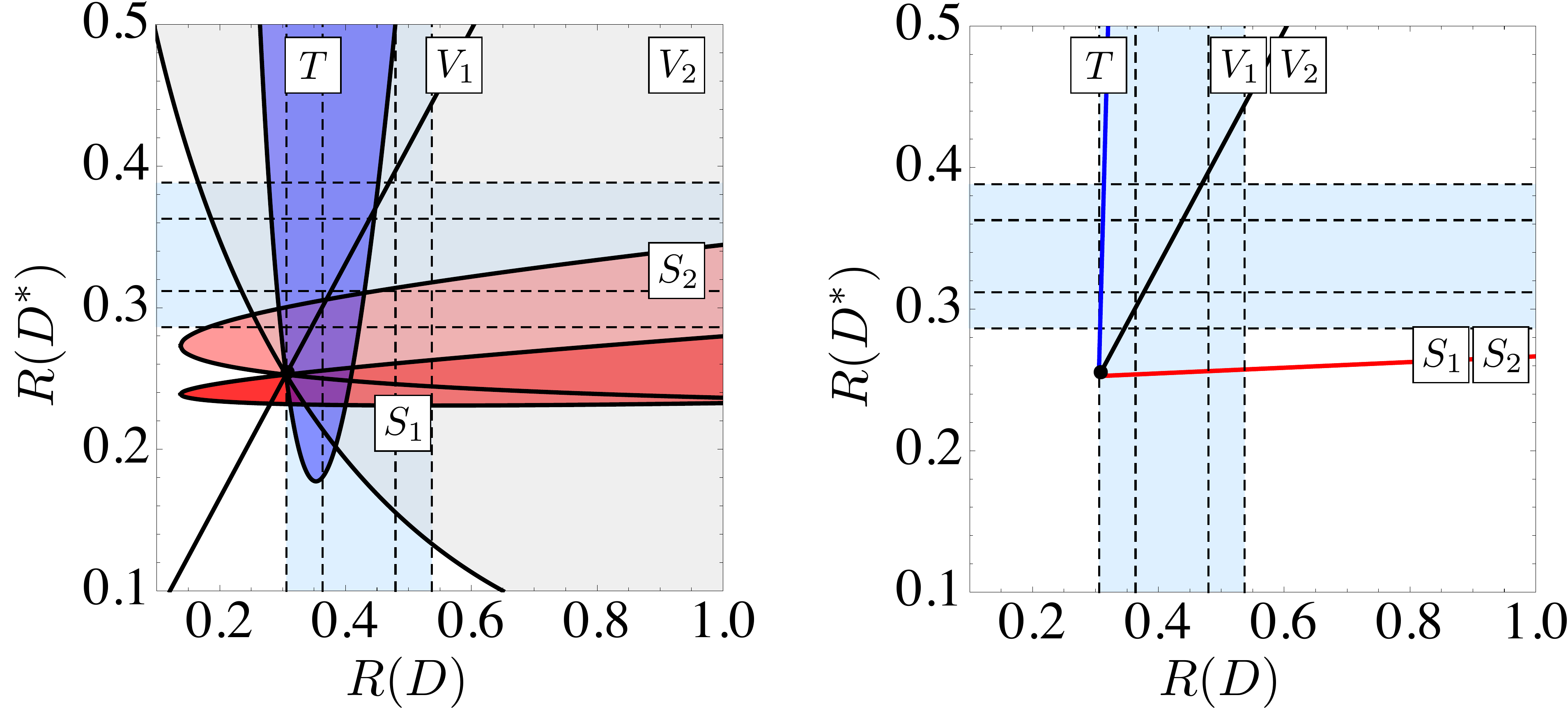}  
\caption{Correlations between $R(D)$ and $R(D^*)$ in the presence of new physics operators of $\mathcal O_X^\tau$ (left) and $\mathcal O_X^{e,\mu}$ (right) for $X=V_{1,2}, S_{1,2}, T$. 
The black dot in each panel indicates the SM prediction. 
The light blue horizontal and vertical bands are the experimental values.}
\label{FIG:DR}
\end{figure}
Since each new physics operator contributes to $R(D)$ and $R(D^*)$ in different ways as seen in Fig.~\ref{FIG:Rratios}, 
it is useful to examine the correlation between $R(D)$ and $R(D^*)$ for the sake of discrimination of new physics operators. 
In the left panel of Fig.~\ref{FIG:DR}, we show $R(D)$ and $R(D^*)$ in the presence of the new physics operators for $\mathcal O_X^\tau$ ($X=V_{1,2}, S_{1,2}$ and $T$). 
The shaded regions are predicted by the indicated operators.  
As is known \cite{TANAKA}, $R(D)$ is more sensitive to the scalar type operators $\mathcal O_{S_1}^\tau$ and $\mathcal O_{S_2}^\tau$ than $R(D^*)$. 
This is due to the angular momentum conservation that gives an extra suppression factor for the vector meson.  
On the other hand,  we find that the tensor type operator $\mathcal O_T^\tau$ exhibits the opposite behavior, that is, $R(D^*)$ is more sensitive to $\mathcal O_T^\tau$ than $R(D)$. 
The vector type operator $\mathcal O_{V_1}^\tau$ gives a unique relation between $R(D)$ and $R(D^*)$, since it is just the SM operator and changes only the overall factor.  
The vector type operator $\mathcal O_{V_2}^\tau$ that contains the right-handed quark current covers a wide region in this plane. 

The correlations between $R(D)$ and $R(D^*)$ for the lepton flavor violating operators $\mathcal O_X^{e,\mu}$ are shown in the right panel of Fig.~\ref{FIG:DR}.  
As these operators do not interfere with the SM one, they always increase $R(D)$ and $R(D^*)$ with different ratios depending on $X$.  

Once a future experiment gives precise values of $R(D)$ and $R(D^*)$, we may exclude some operators depending on the actual experimental values.

\subsection{Correlations between decays rates and polarizations}
\label{CRP}
\begin{figure}
\includegraphics[viewport=5 0 910 580,width=14em]{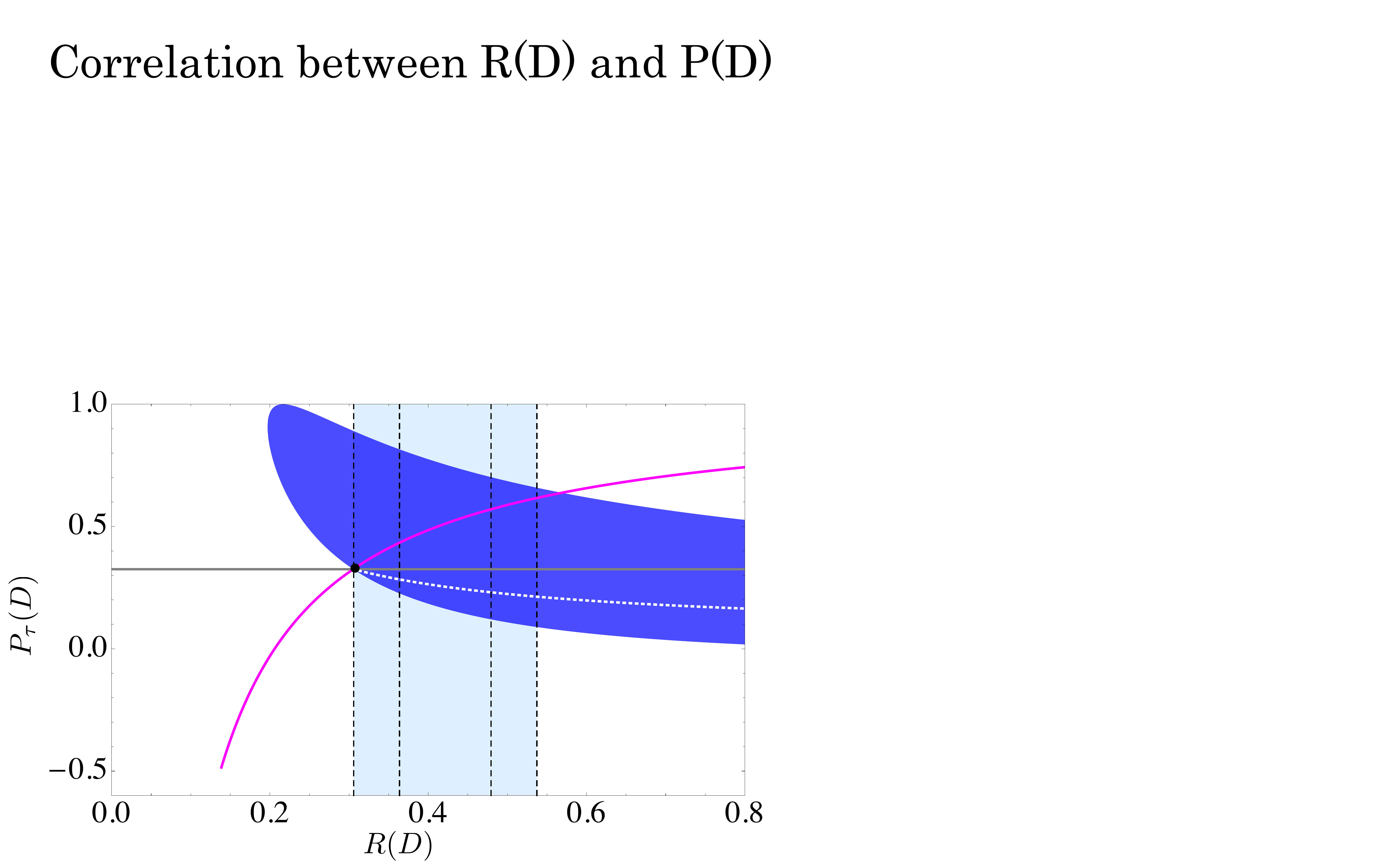}
\includegraphics[viewport=5 0 900 570,width=14em]{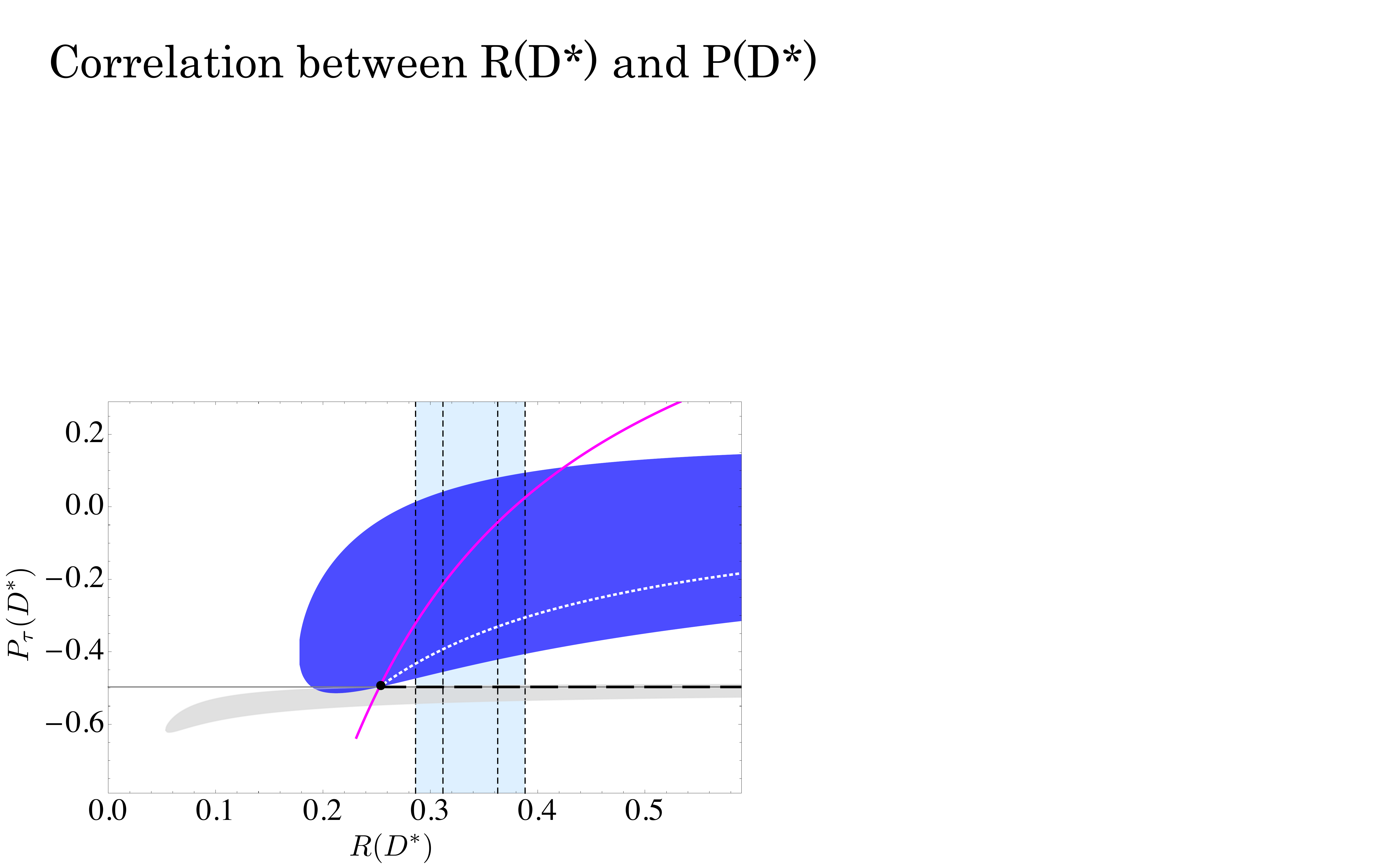}
\includegraphics[viewport=5 0 900 584,width=14em]{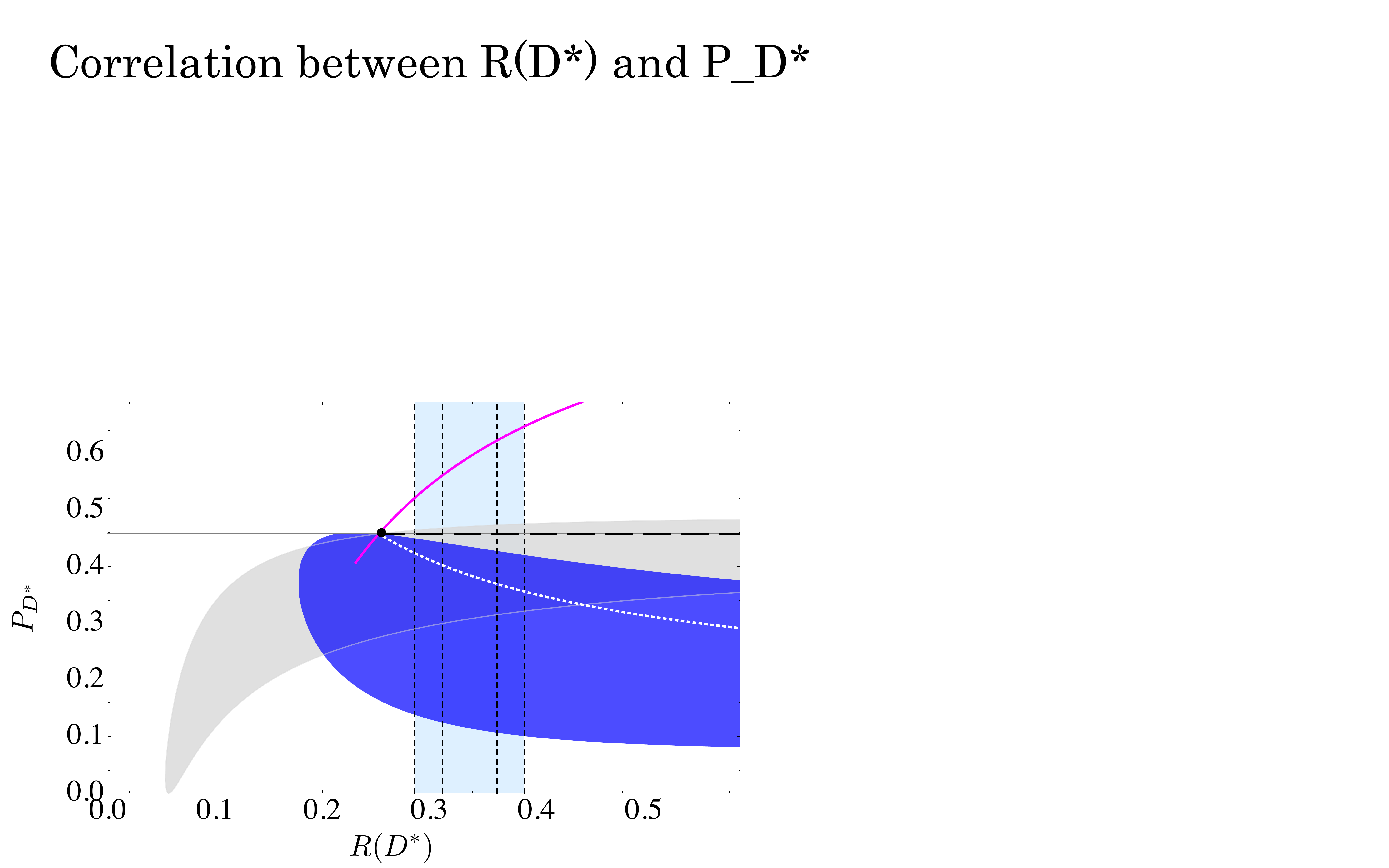}
\caption{Correlations between $R(D^{(*)})$ and $P_\tau(D^{(*)})$, and $R(D^*)$ and $P_{D^*}$ in the presence of new physics operators 
$\mathcal O_{V_1}^{e,\mu,\tau}$ (gray horizontal lines), $\mathcal O_{V_2}^\tau$ (light gray regions), $\mathcal O_{V_2}^{e,\mu}$ (black dashed lines), 
$\mathcal O_{S_{1,2}}^{e,\mu,\tau}$ (magenta curves), $\mathcal O_T^\tau$ (blue regions) and $\mathcal O_T^{e,\mu}$ (white dotted curves).
The black dot in each panel indicates the SM prediction. 
The light blue vertical bands show the experimental constraints.}
\label{FIG:DRandPola}
\end{figure}
As shown in Fig.~\ref{FIG:Pola}, polarizations are also useful observables to identify new physics. 
Here we show correlations between decay rates and polarizations in Fig.~\ref{FIG:DRandPola}. 
The gray horizontal lines represent the correlations for $\mathcal O_{V_1}^{e,\mu,\tau}$, the light gray regions (black dashed lines) for $\mathcal O_{V_2}^{\tau\,(e,\mu)}$, 
the magenta curves for the scalar operators $\mathcal O_{S_{1,2}}^{e,\mu,\tau}$, and the blue regions (white dotted curves) for  $\mathcal O_T^{\tau\,(e,\mu)}$. 
The operator  $\mathcal O_{V_2}^l$ gives the same line as $\mathcal O_{V_1}^l$ in $\bar B \to D\tau\bar\nu$. 
The light blue vertical bands show the experimental constraints on $R(D)$ and $R(D^*)$. 

We find specific features of the scalar operators in this figure.  
The polarizations $P_\tau(D^{(*)})$ and $P_{D^*}$ are uniquely related to the corresponding decay rates in the presence of scalar operators, 
because the scalar operators $\mathcal O_{S_1}^l$ and $\mathcal O_{S_2}^l$ contribute only to $\Gamma^+(D^{(*)})$ and $\Gamma(D^*_L)$. 
The definitions of polarizations may be rewritten as 
\begin{eqnarray}
\label{Eq:PolaScalar1}
 &\ &\left[ \Gamma^+(D^{(*)}) +\Gamma^-(D^{(*)}) \right] \left[1 - P_\tau(D^{(*)})  \right]= 2\Gamma^-(D^{(*)}) \,, \\
\label{Eq:PolaScalar2}
 &\ &\left[ \Gamma(D^*_T) +\Gamma(D^*_L) \right] (1 - P_{D^*})= \Gamma(D^*_T) \,.
\end{eqnarray}
The right-hand sides of these equations are given solely by the SM contributions, and thus the polarizations are definitely determined by the corresponding decay rates as seen in Fig.~\ref{FIG:DRandPola}.
These specific correlations are prominent predictions of the scalar type operators, although we cannot discriminate $\mathcal O_{S_1}^l$ from $\mathcal O_{S_2}^l$ by using these correlations. 

As for the other operators, it is apparent that the SM operator $\mathcal O_{V_1}^\tau$ and $\mathcal O_{V_1}^{e,\mu}$ do not affect the polarizations, 
while the quark right-handed current $\mathcal O_{V_2}^l$ has no effect on the tau polarization in $\bar B\to D\tau\bar\nu$ because the axial vector part does not contribute to this process. 
The operator $\mathcal O_{V_2}^{e,\mu}$ in $\bar B\to D^*\tau\bar\nu$ and the tensor operator $\mathcal O_T^{e,\mu}$ in both the processes predict definite relations between the polarizations and the rates. 
The operator $\mathcal O_{V_2}^\tau$ in $\bar B\to D^*\tau\bar\nu$ and the tensor operator $\mathcal O_T^\tau$ in both the processes have no such specific relations, 
but the covered regions are rather restricted. 

The above correlations that include the polarizations definitely increase the ability to restrict possible new physics. 
We might uniquely identify the new physics operator by these correlations in some cases. 
However their usefulness depends on experimental situations as we will see in the next section.

\section{Model-independent analysis of new physics}
\label{Distinguish}
\begin{figure}
\includegraphics[viewport=0 0 1674 585,width=36em]{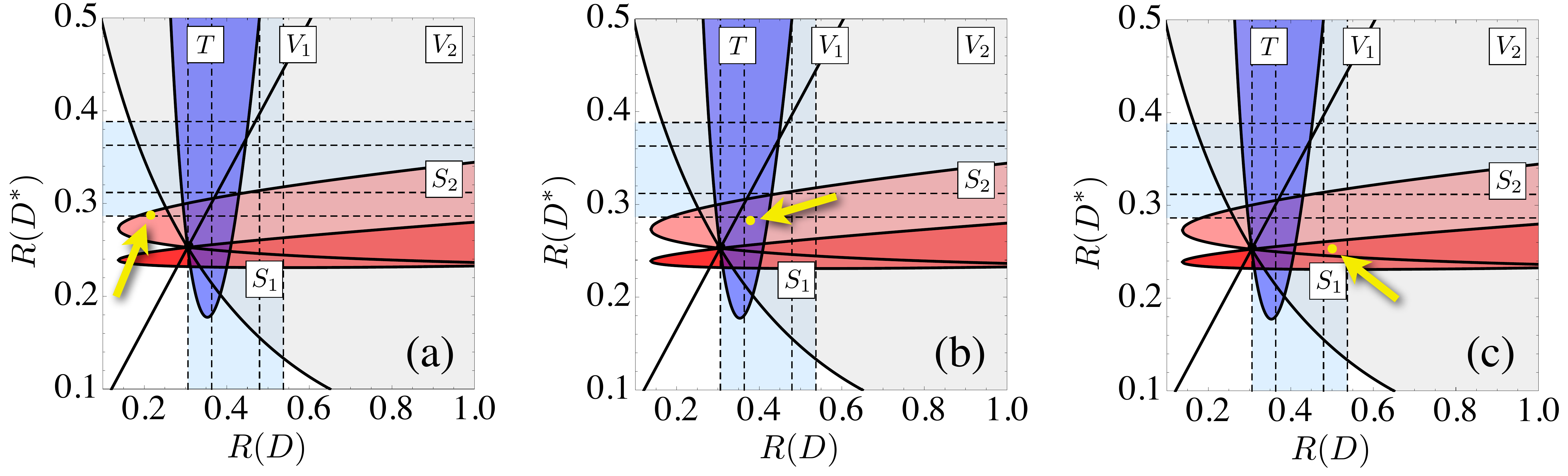}
\caption{Three examples for the discrimination of new physics contribution: 
(a) $R(D)=0.21,\,R(D^*)=0.29$, (b) $R(D)=0.37,\,R(D^*)=0.28$ and (c) $R(D)=0.51,\,R(D^*)=0.25$.}
\label{FIG:example}
\end{figure}
\begingroup
\squeezetable
\begin{table}\begin{center}\begin{tabular}{c|c|ccc|ccc}
 \hline\hline & (a) &  & (b) &  &  & (c) &  \\ 
 \hline $\left( R(D), R(D^*) \right)$ & $(0.21,0.29)$ &  & $(0.37,0.28)$ &  &  & $(0.51,0.25)$ & \\
 \hline X                     & $S_2$                    & $S_2$                    & $V_2$                    & $T$                            &$S_1$                     & $S_2$                  & $V_2$ \\ 
 \hline $C_X^\tau$     & $-1.20\pm i\,0.18$ & $-0.81\pm i\,0.87$ & $0.03\pm i\,0.40$  & $0.16\pm i\,0.14$ & $-0.50\pm i\,1.08$ & $0.21\pm i\,0.56$ & $0.18\pm i0.53$ \\
 \hline $P_\tau(D)$    & $0.02$                   & $0.44$                   & $0.33$                    & $0.22$                       & $0.60$                   & $0.60$                  & $0.33$ \\
 \hline $P_\tau(D^*)$ & $-0.30$                  & $-0.35$                  & $-0.50$                   & $-0.26$                     & $-0.51$                  & $-0.51$                 & $-0.50$ \\
 \hline $P_{D^*}$       & $0.53$                   & $0.51$                   & $0.45$                    & $0.32$                       & $0.45$                   & $0.45$                  & 0.44 \\ \hline\hline
\end{tabular}\end{center}
\caption{Predictions for the polarizations in each of three cases.}\label{Tab:PP}
\end{table}\endgroup
In this section, we illustrate several possibilities to restrict or identify new physics using the observables discussed in the previous section and decay distributions.  
We suppose that $R(D)$ and $R(D^*)$ will be measured more precisely in a future super $B$ factory experiment. 
Then we will determine the Wilson coefficient $C_X^l$ associated with $\mathcal O_X^l$ that is assumed to be dominant except for the SM contribution and predict other observables. 

Here we consider the following three cases of $\left( R(D), R(D^*) \right)$:
\begin{itemize}
  \item[(a)] $(0.21,0.29)$ as shown in Fig.~\ref{FIG:example}(a), in which $\mathcal O_{S_2}^\tau$ is unambiguously identified as the new physics operator. 
  \item[(b)] $(0.37,0.28)$ as shown in Fig.~\ref{FIG:example}(b), in which $\mathcal O_{S_2}^\tau, \mathcal O_{V_2}^\tau$, and $\mathcal O_T^\tau$ are the candidates for the new physics operator. 
  \item[(c)] $(0.51,0.25)$ as shown in Fig.~\ref{FIG:example}(c), in which $\mathcal O_{S_1}^\tau, \mathcal O_{S_2}^\tau$, and $\mathcal O_{V_2}^\tau$ are possible. 
\end{itemize} 
One of $R(D)$ and $R(D^*)$ is chosen to be within the $2\sigma$ range but the other is allowed to deviate more. 
We note that $\mathcal O_X^{e,\mu}$ do not reproduce the assumed sets of $R(D)$ and $R(D^*)$. 
Table~\ref{Tab:PP} summarizes the Wilson coefficient and the predicted polarizations for each case. 

In the case (a), the dominant new physics operator is uniquely determined to be $\mathcal O_{S_2}^\tau$. 
The polarizations can be used to confirm that the deviation from the SM comes from the operator $\mathcal O_{S_2}^\tau$. 
Furthermore the assumption of one-operator dominance will be tested. 

In the case (b), the dominant operator is $\mathcal O_{S_2}^\tau, \mathcal O_{V_2}^\tau$, or $\mathcal O_T^\tau$. 
The predicted values of polarizations vary from operator to operator. 
We will determine the dominant new physics operator by measuring $P_\tau(D^*)$ for example. 
Then the one-operator dominance will be checked by looking up other polarizations. 

\begin{figure}
\includegraphics[viewport=0 0 1172 385,width=36em]{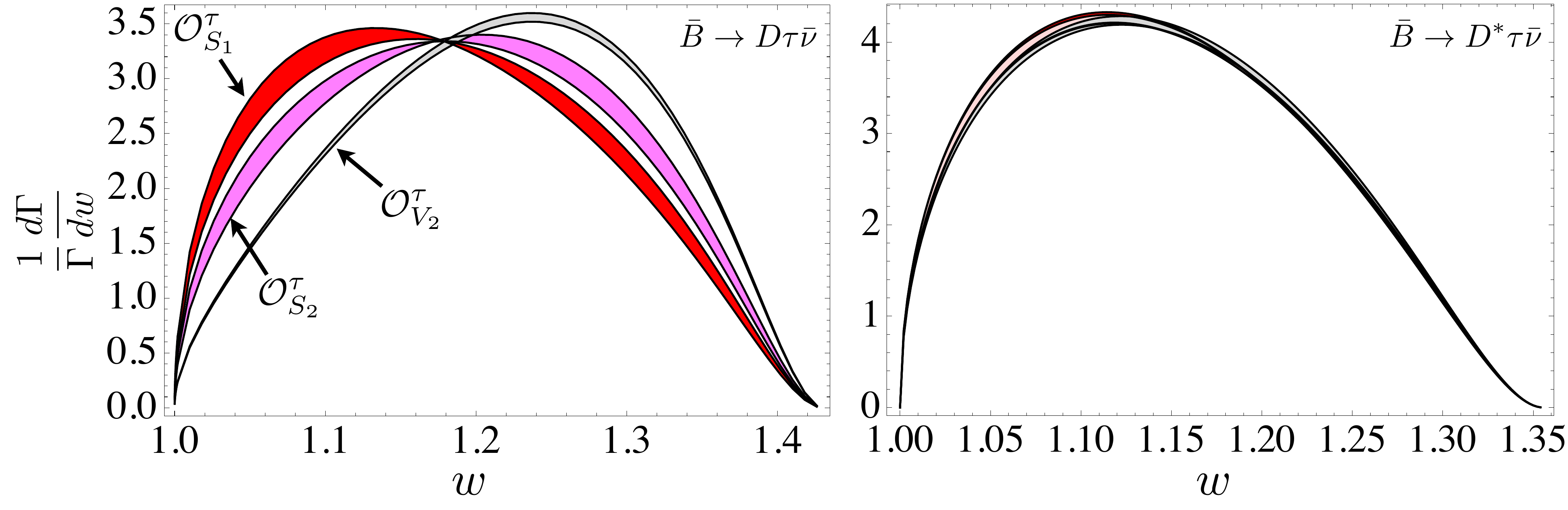}
\caption{Distributions of $w$ in $\bar B\to D\tau\bar\nu$ and $\bar B\to D^*\tau\bar\nu$ for the case (c), where $w=(m_B^2-m_M^2-q^2)/(2m_Bm_M)$. 
The red, magenta, and light gray regions represent the distributions in the cases of $\mathcal O_{S_1}^\tau,\mathcal O_{S_2}^\tau$ and $\mathcal O_{V_2}^\tau$ respectively, 
including the theoretical uncertainties as in Sec.~\ref{Obs}.
The $w$ distributions in $\bar B\to D^*\tau\bar\nu$ are almost identical for the three operators. }
\label{FIG:distribution}
\end{figure}

In the case (c), the dominant operator is $\mathcal O_{S_1}^\tau, \mathcal O_{S_2}^\tau$, or $\mathcal O_{V_2}^\tau$. 
The occurrence of $\mathcal O_{V_2}^\tau$ is distinguished from that of $\mathcal O_{S_1}^\tau$ or $\mathcal O_{S_2}^\tau$ by polarizations. 
Two scalar operators $\mathcal O_{S_1}^\tau$ and $\mathcal O_{S_2}^\tau$, however, predict the same values of the polarizations as explained in Sec.~\ref{CRP}. 
Under such a situation, $q^2$ distributions may discriminate between $\mathcal O_{S_1}^\tau$ and $\mathcal O_{S_2}^\tau$. 
In Fig.~\ref{FIG:distribution}, we present $q^2$ distributions for $\mathcal O_{S_1}^\tau$, $\mathcal O_{S_2}^\tau$, and $\mathcal O_{V_2}^\tau$ in the case (c). 
We note that the abscissa is $w=(m_B^2-m_M^2-q^2)/(2m_Bm_M)$ instead of $q^2$. 
The $q^2$ distribution in $\bar B\to D\tau\bar\nu$ turns out to be useful for the discrimination in this case. 

\section{Model analysis}
\label{Model}
In this section, we discuss some new physics models which affect $\bar B \to D^{(*)} \tau \bar\nu$ based on the results in Sec.~\ref{PredandConst}.

\subsection{2HDMs and MSSM}
\begin{table}\begin{center}\begin{tabular}{c|cccc}
 \hline 
 \hline              & Type\,I             & Type\,II            & Type\,X & Type\,Y \\
 \hline $\xi_d$ & $\cot^2 \beta$  & $\tan^2 \beta$ & $-1$       & $-1$ \\
 \hline $\xi_u$ & $-\cot^2 \beta$ & $1$                  & $1$        & $-\cot^2 \beta$ \\
 \hline 
 \hline
\end{tabular}\end{center}
\caption{Parameters $\xi_{d,u}$ in each type of 2HDMs.}\label{Tab:2HDM}
\end{table}
As known well, the charged Higgs boson in two Higgs doublet models (2HDMs) contributes to $\bar B \to D^{(*)} \tau \bar\nu$ and its effect is enhanced in some cases. 
In order to forbid flavor changing neutral currents (FCNC) at the tree level, a $Z_2$ symmetry is often imposed in this class of models and it results in four distinct 2HDMs
\cite{THDMBarger, THDMGrossman, THDMother, THDMKanemura} . 
Their Yukawa terms are described as 
\begin{eqnarray}
 \mathcal{L}_\text{Y} &=& -\bar Q_L Y_u \tilde H_2 u_R -\bar Q_L Y_d H_2 d_R -\bar L_L Y_\ell H_2 \ell_R +\text{h.c.}\quad \text{(type I)}  \,, \\
 \mathcal{L}_\text{Y} &=& -\bar Q_L Y_u \tilde H_2 u_R -\bar Q_L Y_d H_1 d_R -\bar L_L Y_\ell H_1 \ell_R +\text{h.c.}\quad \text{(type II)}  \,, \\
 \mathcal{L}_\text{Y} &=& -\bar Q_L Y_u \tilde H_2 u_R -\bar Q_L Y_d H_2 d_R -\bar L_L Y_\ell H_1 \ell_R +\text{h.c.}\quad \text{(type X)}  \,, \\
 \mathcal{L}_\text{Y} &=& -\bar Q_L Y_u \tilde H_2 u_R -\bar Q_L Y_d H_1 d_R -\bar L_L Y_\ell H_2 \ell_R +\text{h.c.}\quad \text{(type Y)}  \,,
\end{eqnarray}
where $H_{1,2}$ are Higgs doublets defined as 
\begin{eqnarray}
 H_i = \begin{pmatrix} h_i^+ \\ (v_i +h_i^0)/\sqrt 2 \end{pmatrix}, \quad \tilde H_i = i\sigma_2 H_i, 
\end{eqnarray}
and $v_i$ denotes the vacuum expectation value (VEV) of $H_i$. 
The ratio of two VEVs is defined as $\tan \beta = v_2/v_1$ and $v=\sqrt{v_1^2+v_2^2}=246\text{GeV}$.
In type I, all masses of quarks and leptons are given by the VEV of $H_2$. 
In type II, the up-type quarks obtain their masses from $H_2$, while the down-type quarks and leptons from $H_1$. 
In type X, $H_2$ and $H_1$ are responsible to the quark and lepton masses respectively. 
The masses of the down-type quarks are given by $H_1$ and other fermions acquire their masses from $H_2$ in type Y.  
Under this definition $v_2$ generates up-quark masses in any type of Yukawa interaction.  

These 2HDMs contain a pair of physical charged Higgs bosons, which contributes to $\bar B \to D^{(*)} \tau \bar\nu$. 
The relevant Wilson coefficients introduced in Eq.~(\ref{Eq:EL}) are given by 
\begin{eqnarray}
 C_{S_1}^\tau = - \frac{m_b m_\tau}{m_{H^\pm}^2} \xi_d \,,\quad C_{S_2}^\tau &=& - \frac{m_c m_\tau}{m_{H^\pm}^2} \xi_u \,.
\end{eqnarray}
where $m_{H^\pm}$ is the mass of the charged Higgs boson. 
The parameters $\xi_d$ and $\xi_u$ are presented in Table~\ref{Tab:2HDM}.  
To have a sizable charged Higgs effect, $|\xi_{d,u}|$ should be much larger than unity taking the experimental bound on the charged Higgs mass into account. 
Then the case of $\xi_u=1$ or $\xi_d=-1$ is not acceptable. 
The case of $\xi_u=-\cot^2\beta$ or $\xi_d=\cot^2\beta$ with $\cot^2\beta\gg1$ is unnatural since the top Yukawa interaction becomes nonperturbative.  
The requirement for the top Yukawa interaction to be perturbative results in $\tan\beta \gtrsim 0.4$ \cite{THDMKanemura}. 
Therefore, $\xi_d =\tan^2\beta$ in 2HDM of type II is suitable and only $C_{S_1}^\tau$ is enhanced. 
As we have shown in Sec.~\ref{PredandConst}, however, it is difficult to explain the current experimental results by $\mathcal O_{S_1}^\tau$ alone.  
We find that this model is disfavored with about $4\sigma$ using the combined experimental values in Eq.~(\ref{Eq:combined}).
This result is consistent with the recent study in Ref.~\cite{Babar2012}.

\begin{table}\begin{center}\begin{tabular}{c|ccc}
 \hline 
 \hline              & $Z_u$            & $Z_d$            & $Z_\ell$  \\
 \hline Type\,I  & $\frac{\sqrt2 M_u}{v} \cot\beta -\epsilon_u \sin\beta(1+\cot^2\beta) $  & $-\frac{\sqrt2 M_d}{v} \cot\beta +\epsilon_d \sin\beta(1+\cot^2\beta) $ & $-\frac{\sqrt2 M_\ell}{v}\cot\beta$ \\
 \hline Type\,II & $\frac{\sqrt2 M_u}{v} \cot\beta -\epsilon_u \cos\beta(\tan\beta+\cot\beta) $ & $\frac{\sqrt2 M_d}{v} \tan\beta -\epsilon_d \sin\beta(\tan\beta+\cot\beta) $ & $\frac{\sqrt2 M_\ell}{v}\tan\beta$ \\
 \hline Type\,X & $\frac{\sqrt2 M_u}{v} \cot\beta -\epsilon_u \sin\beta(1+\cot^2\beta) $& $-\frac{\sqrt2 M_d}{v} \cot\beta +\epsilon_d \sin\beta(1+\cot^2\beta) $ & $\frac{\sqrt2 M_\ell}{v}\tan\beta$ \\
 \hline Type\,Y & $\frac{\sqrt2 M_u}{v} \cot\beta -\epsilon_u \cos\beta(\tan\beta+\cot\beta) $ & $\frac{\sqrt2 M_d}{v} \tan\beta -\epsilon_d \sin\beta(\tan\beta+\cot\beta) $ &$-\frac{\sqrt2 M_\ell}{v}\cot\beta$ \\
 \hline
 \hline
\end{tabular}\end{center}
\caption{The matrices $Z_{u,d,\ell}$ in each type of the 2HDM in the presence of the $Z_2$ breaking terms.}\label{Tab:general2HDM}
\end{table}
A possible solution within 2HDMs is to violate the $Z_2$ symmetry at the cost of FCNC. 
We introduce the following $Z_2$ breaking terms in the above four models: 
\begin{eqnarray}
 \Delta \mathcal{L}_\text{Y} &=& -\bar Q_L \epsilon'_u \tilde H_1 u_R -\bar Q_L \epsilon'_d H_1 d_R +\text{h.c.}\quad \text{(for types I and X)} \,, \\
 \Delta \mathcal{L}_\text{Y} &=& -\bar Q_L \epsilon'_u \tilde H_1 u_R -\bar Q_L \epsilon'_d H_2 d_R +\text{h.c.}\quad \text{(for types II and Y)} \,, \label{Eq:Z2type2}  
\end{eqnarray}
where $\epsilon'_{u,d}$ are $3\times3$ matrices that control FCNC and the quark fields are those in the weak basis.  
Writing the Yukawa terms $\mathcal{L}_\text{Y} +\Delta \mathcal{L}_\text{Y} $ in terms of mass eigenstates, we obtain the following physical charged Higgs interaction terms: 
\begin{eqnarray}
 \mathcal{L}_{H^\pm} &=&\left( \bar u_R Z_u^\dag V_\text{CKM} d_L +\bar u_L V_\text{CKM} Z_d d_R +\bar\nu_L Z_\ell \ell_R \right) H^+ +\text{h.c.} \,, 
\end{eqnarray}
where $V_\text{CKM}$ is Cabibbo-Kobayashi-Maskawa (CKM) matrix. 
Table~\ref{Tab:general2HDM} shows the expressions of $Z_{u,d,\ell}$, 
where $M_{u,d,\ell}$ denote the diagonal up-type quark, down-type quark, and lepton mass matrices, 
and $\epsilon_{u,d}$ represent matrices $\epsilon'_{u,d}$ in the quark mass basis. 

The FCNC in the down-quark sector is strongly constrained, so that $\epsilon_d$ is negligible in the present analysis. 
On the other hand, constraints on the FCNC in the up quark sector are rather weak. 
Recently the 2HDM of type II that allows FCNC in the up quark sector is studied to explain $\bar B \to D^{(*)} \tau \bar\nu$ \cite{CGK}. 
Table~\ref{Tab:general2HDM} implies that the operator $\mathcal O_{S_2}^\tau$ in types II and X might be significant for large $\tan\beta$. 
Then the corresponding Wilson coefficient is given by 
\begin{eqnarray}
 C_{S_2}^\tau \simeq \frac{V_{tb}}{\sqrt2 V_{cb}} \frac{v m_\tau}{m_{H^\pm}^2} (\epsilon_u^{tc})^* \sin\beta\tan\beta. 
\end{eqnarray} 
As seen in Fig.~\ref{FIG:Contour} the current experimental results are described by the 2HDM of the type II or X with FCNC provided that $|\epsilon_u^{tc}| \sim 1$. 
If this is the case, we expect sizable deviations in polarizations $P_\tau(D^{(*)})$ and $P_{D^*}$ from the SM as designated by the magenta curves in Fig.~\ref{FIG:DRandPola}. 

The charged Higgs effects on $\bar B \to D^{(*)} \tau \bar\nu$ in the minimal supersymmetric standard model (MSSM) are the same as those in the 2HDM of type\,II at the tree level. 
Loop corrections induce non-holomorphic terms $\epsilon_{u,d}$ in Eq.~(\ref{Eq:Z2type2}) \cite{IKO,Crivellin}. 
However it seems difficult to enhance $\epsilon_u^{tc}$ to be $O(1)$.  
Thus the sufficient enhancement of $\mathcal O_{S_2}^\tau$ is unlikely in MSSM.

\subsection{MSSM with R-parity violation}
The $R$-parity violating (RPV) MSSM \cite{RPVrev2} may also affect $\bar B \to D^{(*)} \tau \bar\nu$ \cite{RPVbound2,RPVrev,DM}. 
We consider the following superpotential:
\begin{equation}
 W_\text{RPV} = \frac{1}{2} \lambda_{ijk} L_i L_j E^c_k +\lambda'_{ijk} L_i Q_j D^c_k \,, 
\end{equation}
where $\lambda_{ijk}$ and $\lambda'_{ijk}$ are RPV couplings and $i,j,k$ are generation indices. 
Apart from the charged Higgs contribution, there are two kinds of diagrams which contribute to $\bar B \to D^{(*)} \tau \bar\nu$, that is, the slepton and down squark exchanging diagrams. 
The corresponding effective Lagrangian is written as 
\begin{equation}\label{Eq:RPV}
 \mathcal{L}_\text{eff}^\text{RPV} 
 =-\sum_{j,k=1}^3 V_{2k} \left[ \frac{ \lambda_{ij3} \lambda'^*_{jk3}}{m^2_{\tilde l_L^j}}\, \bar c_L b_R\, \bar \tau_R \nu_L^i
   +\frac{ \lambda'_{i3j} \lambda'^*_{3kj}}{m^2_{\tilde d_R^j}}\, \bar c_L \left( \tau^c \right)_R\, \left( \bar \nu^c \right)_R^i b_L \right]\,,
 \end{equation}
where $m_{\tilde l_L^j}\, (m_{\tilde d_R^j})$ is the mass of the slepton (down squark) for the $j$-th generation and $V_{ij}$ is the component of CKM matrix. 
Here we assume that the slepton and down squark mass matrices are diagonal in the super-CKM basis for simplicity. 
Using Fierz identity the second term in Eq.~(\ref{Eq:RPV}) is rewritten as  
\begin{equation}
 \bar c_L \left( \tau^c \right)_R\, \left( \bar \nu^c \right)_R^i b_L = \frac{1}{2} \bar c_L \gamma^\mu b_L\, \bar \tau_L \gamma_\mu \nu_L^i. 
\end{equation}
Then, the corresponding coefficients, defined in Eq.~(\ref{Eq:EL}), are expressed as
\begin{eqnarray}
 C_{S_1}^{l_i} =\frac{1}{2\sqrt2 G_F V_{cb}} \sum_{j,k=1}^3 V_{2k} \frac{ \lambda_{ij3} \lambda'^*_{jk3}}{m^2_{\tilde l_L^j}} \,, \quad
 C_{V_1}^{l_i} =\frac{1}{2\sqrt2 G_F V_{cb}} \sum_{j,k=1}^3 V_{2k} \frac{ \lambda'_{i3j} \lambda'^*_{3kj}}{2m^2_{\tilde d_R^j}} \,, \label{Eq:RPVcoupling}
\end{eqnarray}
where $l_1=e,\,l_2=\mu,$ and $l_3=\tau$ represent the flavor of the neutrino.  

The case that $C_{S_1}^{l_i}$ is dominant contribution is disfavored as discussed in Sec.~\ref{PredandConst}. 
On the other hand $C_{V_1}^{l_i}$ has an allowed region as shown in Fig.~\ref{FIG:Contour}. 
We find that $\left |C_{V_1}^\tau \right |>0.08$ and $\left |C_{V_1}^{e,\mu} \right |>0.42$ (90\% C.L.) are required to explain the current experimental results. 
However the same products of RPV couplings as in $C_{V_1}^{l_i}$'s also contribute to $B \to X_s \nu\bar\nu$ \cite{RPVbound1}.  
The upper limit of the branching ratio, $\mathcal B(B \to X_s \nu\bar\nu) <6.4 \times 10^{-4}$ (90\% C.L.) \cite{ALEPH}, puts a strong constraint on RPV couplings. 
It turns out that $\left | C_{V_1}^{l_i} \right | <0.03$ is imposed, which contradicts the above requirement for $\bar B \to D^{(*)} \tau \bar\nu$. 
Thus MSSM with RPV is not likely to be consistent with both $\bar B \to D^{(*)} \tau \bar\nu$ and $B \to X_s \nu\bar\nu$ at the same time.

\subsection{Leptoquark models}\label{LQ}
There are ten possible leptoquark models that respect the symmetry of the SM \cite{LQBRW}. 
Among them, the following leptoquark model is interesting in the sense that the tensor operator is generated \cite{JPL}, 
\begin{equation}\label{Eq:LQmodel}
 \mathcal L_\text{LQ} = \left( \lambda_{ij} \bar Q^i \varepsilon e_R^j +\lambda'_{ij} \bar u_R^i L^j \right) S_\text{LQ} \,, 
\end{equation}
where $S_\text{LQ}$ is the scalar leptoquark with $SU(3)\times SU(2)\times U(1)_Y$ quantum number being $(3,2,7/6)$. 
The effective Lagrangian is represented as 
\begin{equation}
 -\mathcal{L}_\text{eff}^\text{LQ} 
 =\frac{\lambda_{33}^* \lambda'_{2i}}{m_{S_\text{LQ}}^2} \left( \bar \tau_R b_L \right) \left(\bar c_R \nu_L^i \right)
 =-\frac{\lambda_{33}^* \lambda'_{2i}}{2m_{S_\text{LQ}}^2} \left( \bar c_R b_L \bar\tau_R \nu_L^i +\frac{1}{4}\bar c_R \sigma^{\mu\nu} b_L \bar\tau_R \sigma_{\mu\nu} \nu_L^i  \right) . 
 \end{equation}
Both $\mathcal O_{S_2}^{l_i}$ and $\mathcal O_T^{l_i}$ appear at the same time, however the Wilson coefficients are related to each other: 
\begin{eqnarray}\label{Eq:LQresult}
 C_{S_2}^{l_i} =4C_T^{l_i} =\frac{-1}{2\sqrt2 G_F V_{cb}} \cdot \frac{ \lambda_{33}^* \lambda'_{2i}}{2m_{S_\text{LQ}}^2}
\simeq -0.6 \left(\frac{\lambda_{33}^* \lambda'_{2i}}{0.4}\right)  \left(\frac{500\,\text{GeV}}{m_{S_\text{LQ}} }\right)^2  \,. 
\end{eqnarray}
Thus, a similar analysis as the case of one dominant operator can be made.  

The bottom right panel in Fig.~\ref{FIG:Contour} shows the constraint on the above Wilson coefficient $C_{S_2}^\tau (=4C_T^\tau)$. 
We find that the present experimental results are explained with $C_{S_2}^\tau \simeq \pm i\, 0.6$ or $|C_{S_2}^{e,\mu}| \simeq 0.6$. 
The relevant mass scale of the leptoquark could be  $\sim500$ GeV as seen in Eq.~(\ref{Eq:LQresult}).  
\begin{figure}
\includegraphics[viewport=5 0 910 580,width=14em]{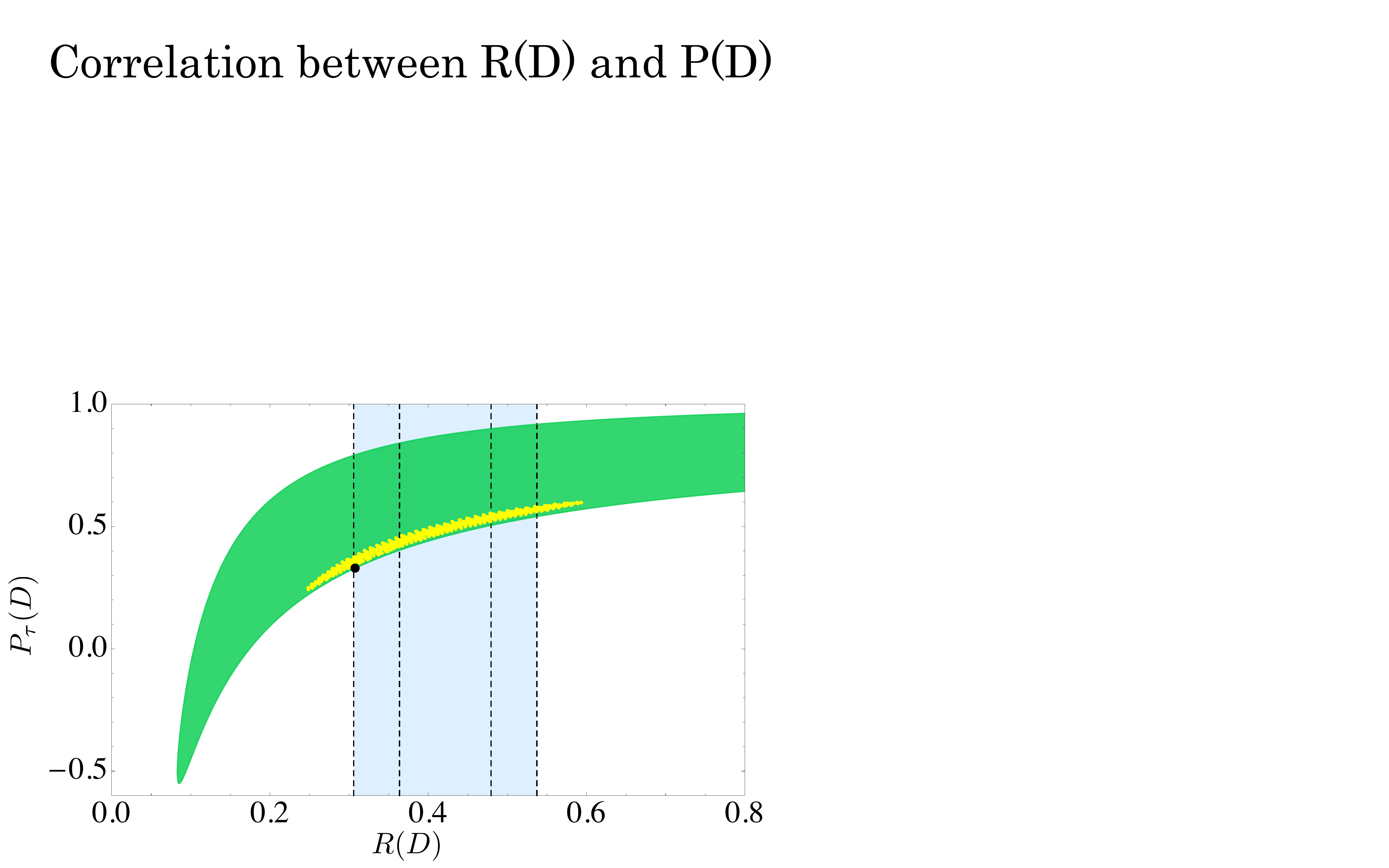}
\includegraphics[viewport=5 0 900 570,width=14em]{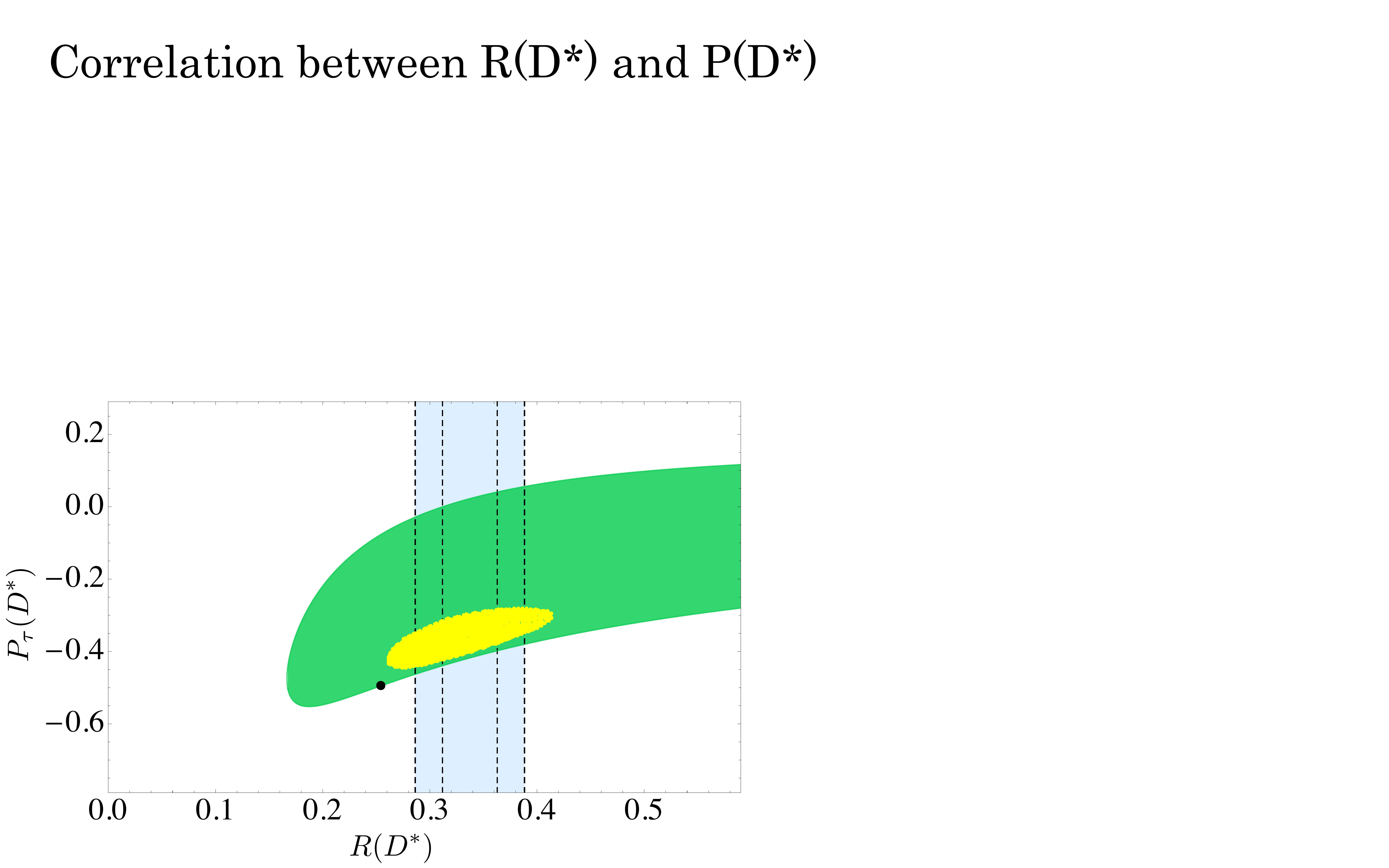}
\includegraphics[viewport=5 0 900 584,width=14em]{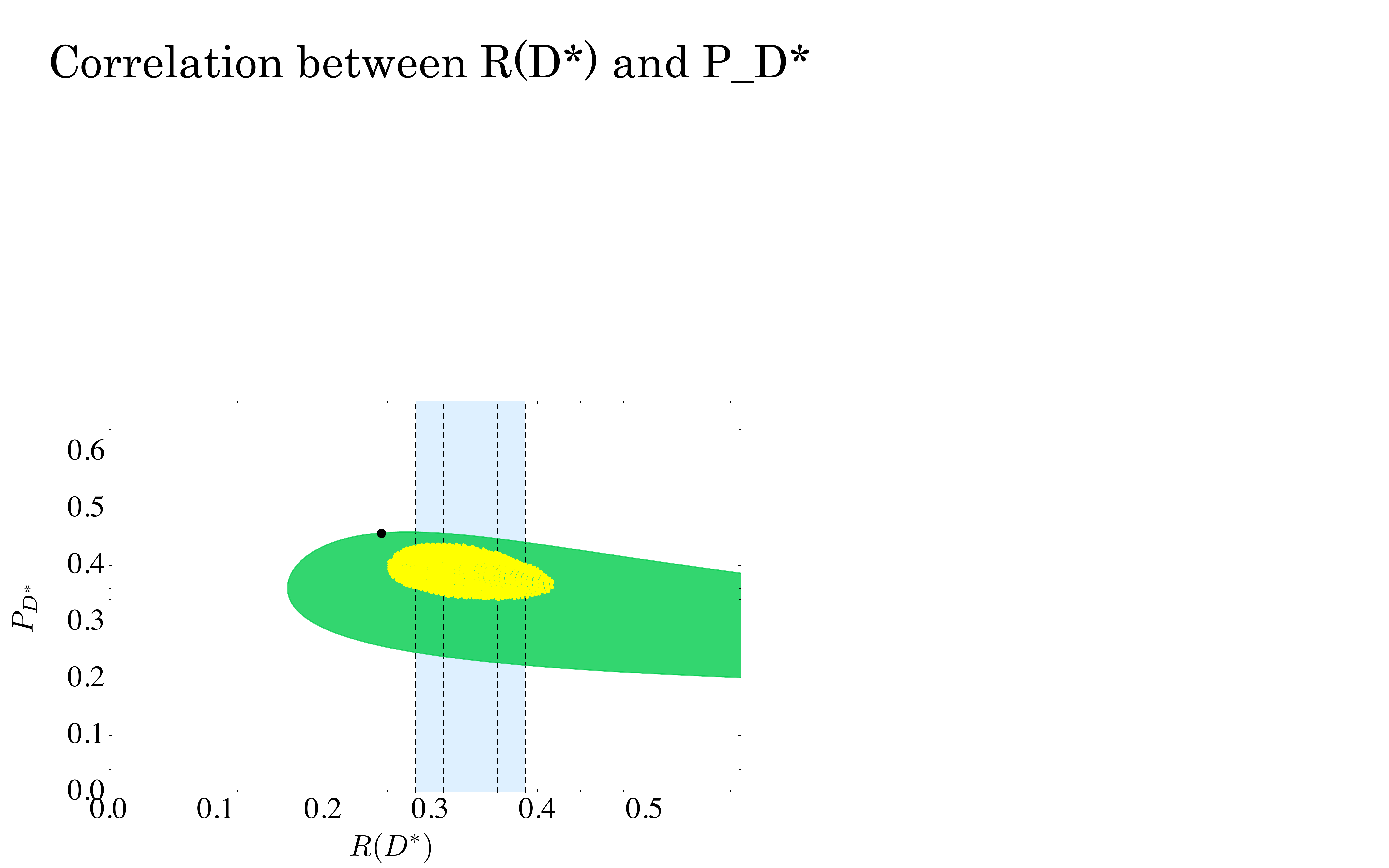}
\caption{Correlations between $R(D^{(*)})$ and $P_\tau(D^{(*)})$, and $R(D^*)$ and $P_{D^*}$ in the leptoquark model of Eq.~(\ref{Eq:LQmodel}) are represented by green regions. 
Yellow regions indicate the constraints from both the present experimental bounds on $R(D)$ and $R(D^*)$. 
The black dot in each panel stands for the SM prediction. }
\label{FIG:DRandPolaLQ}
\end{figure}
In Fig.~\ref{FIG:DRandPolaLQ}, we show the correlations between decay rates and polarizations in the above leptoquark model. 
The green regions represent the correlations, $R(D^{(*)})$--$P_\tau(D^{(*)})$ and $R(D^*)$--$P_{D^*}$, as in Fig.~\ref{FIG:DRandPola}. 
Taking into account both the experimental constraints on $R(D)$ and $R(D^*)$ at the same time, we find the present allowed regions at 99\% C.L. as shown by the yellow regions.

\section{Conclusions} 
\label{Conclusion}
We have studied the exclusive semi-tauonic $B$ decays, 
$\bar B\to D^{(*)}\tau\bar\nu$, in the model-independent manner based on 
the effective Lagrangian including all the possible four-Fermi operators.
It has turned out that the current experimental values of $R(D)$ and $R(D^*)$ are not explained by the operator $O_{S_1}^l$ nor $O_{S_2}^{e,\mu}$ alone, 
while the other operators $O_{S_2}^\tau$, $O_{V_1}^l$, $O_{V_2}^l$, and $O_T^l$ reasonably work under the assumption of one-operator dominance. 
More precise data that will be given in a future super $B$ factory experiment
will allow us to identify the relevant new physics operator among these operators if the deviation from the SM persists. 
We have pointed out that correlations among observables including the longitudinal tau polarizations, the $D^*$ polarization and $q^2$ distributions are useful in distinguishing among possible new physics. 
This observation for the most general effective Lagrangian in Eq.~(\ref{Eq:EL}) confirms and extends results in the literature for limited new physics cases \cite{FKN,DDG,CJLP,OUR}. 

Furthermore, we have studied several interesting models that may describe the deviations of $R(D)$ and $R(D^*)$ from the SM based on our model-independent analysis. 
In 2HDMs without tree-level FCNC and MSSM, only $O_{S_1}^\tau$ could be enhanced and thus are unlikely to explain the deviations, 
while $O_{S_2}^\tau$ generated in 2HDMs of type II and type X with FCNC might be sizable depending on the magnitude of the $t\to c$ FCNC parameter.
The parameter region of MSSM with $R$-parity violation that explains the deviations in $R(D)$ and $R(D^*)$ has turned out to be inconsistent with the experimental bound on $B\to X_s\nu\bar\nu$. 
The scalar leptoquark model that simultaneously induces $O_{S_2}^l$ and $O_T^l$ describes the experimental data. 

In conclusion, $\bar B\to D^{(*)}\tau\bar\nu$ is a powerful tool to explore new physics in the charged current. 
It is important to study combinations of observables including particle polarizations and $q^2$ distributions as well as the decay rates in order to clarify possible new physics 
in a model-independent manner. 
The model-independent analysis gives a firm footing in examining models of new physics.

\begin{acknowledgements}
The authors thank A.~Tayduganov for his useful comments on the manuscript. 
This work is supported in part by the Grant-in-Aid for Science Research, Ministry of Education, Culture, Sports, Science and Technology, Japan, under Grants No.~20244037 (M.T.) and No.~248920 (R.W.). 
\end{acknowledgements}

\appendix

\section{Leptonic amplitudes}
\label{Ap:LA}
In this appendix, we summarize the expressions of the leptonic amplitudes. 
The vector type leptonic amplitudes are given by 
\begin{eqnarray}
 L^+_\pm &=& \pm\sqrt{2} m_\tau v \sin \theta_\tau \ ,\\
 L^+_0 &=& 2 m_\tau v \cos \theta_\tau \ ,\\
 L^+_s &=& -2 m_\tau v\ ,\\
 L^-_\pm &=& \sqrt{2} \sqrt{q^2} v(1\pm \cos \theta_\tau)\ ,\\
 L^-_0 &=& -2 \sqrt{q^2} v \sin \theta_\tau \ ,\\
 L^-_s &=& 0\ ,
\end{eqnarray}
where $v =\sqrt{1-m^2_\tau /q^2}$. 
The scalar type leptonic amplitudes are written as
\begin{eqnarray}
 L^+&=& -2 \sqrt{q^2} v\ , \\
 L^-&=&0\ .
\end{eqnarray}
The tensor type leptonic amplitudes are 
\begin{eqnarray}
 L^\pm_{\lambda \lambda} &=& 0\ , \\
 L^+_{0\pm}&=&-L^+_{\pm0}= -\sqrt{2} \sqrt{q^2} v \sin \theta_\tau \ ,\\
 L^+_{+-}&=&-L^+_{-+}=L^+_{s0}=-L^+_{0s}= 2 \sqrt{q^2} v \cos \theta_\tau \ ,\\
 L^+_{\pm s}&=&-L^+_{s \pm}= \mp \sqrt{2} \sqrt{q^2} v \sin \theta_\tau \ ,\\
 L^-_{0\pm}&=&-L^-_{\pm0}= \mp \sqrt{2} m_\tau v (1\pm \cos \theta_\tau) \ ,\\
 L^-_{+-}&=&-L^-_{-+}=L^-_{s0}=-L^-_{0s}= - 2 m_\tau v \sin \theta_\tau \ , \\
 L^-_{\pm s}&=&-L^-_{s \pm}= -\sqrt{2} m_\tau v (1\pm \cos \theta_\tau)\ .
\end{eqnarray}
The subscript $l$ that represents the neutrino flavor is omitted for brevity. 

\section{Form factors and hadronic amplitudes}
\label{Ap:HA}
Here, we show the definitions of the relevant  form factors and the hadronic amplitudes  in $\bar B \to D^{(*)} \tau \bar\nu$. 
We also discuss how to evaluate the scalar and tensor form factors by using the equations of motion.

\subsection{Vector and axial vector operators}
We define the form factors of the vector and axial vector operators in $\bar B \to D^{(*)} \tau \bar\nu$ as, 
\begin{eqnarray}
\label{Eq:vectorD}
 \langle D(p_D)| \bar c\gamma^\mu b |\bar B(p_B)\rangle &=& \sqrt{m_Bm_D} \left [ h_+(w) (v+v')^\mu + h_-(w) (v-v')^\mu \right ] \,, \\
\label{Eq:vectorDstar}
 \langle D^*(p_{D^*},\epsilon)| \bar c \gamma^\mu b| \bar B (p_B) \rangle &=& i \sqrt{m_Bm_{D^*}}\, h_V(w) \varepsilon^{\mu\nu\rho\sigma} \epsilon^*_\nu v'_\rho v_\sigma \,, \\
\label{Eq:axialvectorDstar}
 \langle D^*(p_{D^*},\epsilon)| \bar c \gamma^\mu \gamma^5 b| \bar B (p_B) \rangle
 &=& \sqrt{m_Bm_{D^*}}\, \big [ h_{A_1}(w) (w+1) \epsilon^{*\mu}\notag \\
 &\ &-(\epsilon^* \cdot v) \left(h_{A_2}(w) v^\mu +h_{A_3}(w) v'^\mu \right) \big ],
\end{eqnarray}
where $v=p_B/m_B, v'=p_M/m_M$, and $w=v\cdot v'$ in $\bar B \to M \tau\bar\nu$. 
The polarization vector of $D^*$ is indicated as $\epsilon^\mu$. 

Using these definitions, the hadronic amplitudes defined in Eqs.~(\ref{Eq:HAV1}) and (\ref{Eq:HAV2}) are represented as
\begin{eqnarray}
 H_{V_1,\pm}^s &=& H_{V_2,\pm}^s =0\,,\\
 H_{V_1,0}^s &=& H_{V_2,0}^s = m_B \sqrt{r} \frac{\sqrt{w^2-1}}{\sqrt{\hat q^2(w)}} \left[ (1+r)h_+(w) -(1-r)h_-(w) \right]\,,\\
 H_{V_1,s}^s &=& H_{V_2,s}^s =m_B \sqrt{r} \frac{1}{\sqrt{\hat q^2(w)}} \left[ (1-r)(w+1)h_+(w) -(1+r)(w-1)h_-(w) \right]\,, 
\end{eqnarray}
in $\bar B \to D \tau \bar\nu$ and those in $\bar B \to D^* \tau \bar\nu$ are given by 
\begin{eqnarray}
 H_{V_1,\pm}^\pm 
 &=& -H_{V_2,\mp}^\mp \notag \\
 &=& m_B \sqrt{r} \left [ (w+1)h_{A_1}(w) \mp \sqrt{w^2-1}h_V(w) \right ] \,,\\
 H_{V_1,0}^0 
 &=& -H_{V_2,0}^0 \notag \\
 &=& m_B \sqrt{r} \frac{w+1}{\sqrt{\hat q^2(w)}} \left [ -(w-r)h_{A_1}(w) +(w-1) (r h_{A_2}(w) +h_{A_3}(w)) \right ],\\
 H_{V_1,s}^0 
 &=& -H_{V_2,s}^0 \notag \\
 &=& m_B \sqrt{r} \frac{\sqrt{w^2-1}}{\sqrt{\hat q^2(w)}} \left [ -(w+1)h_{A_1}(w) +(1-rw) h_{A_2}(w) +(w-r) h_{A_3}(w) \right ],\\
 \text{others} &=& 0.
\end{eqnarray}
In the heavy quark limit, we obtain $h_+(w)=h_{A_{1,3}}(w)=h_V(w)=\xi(w)$ and $h_-(w)=h_{A_{2}}(w)=0$ where $\xi(w)$ is Isgur-Wise function \cite{IW}. 
In order to evaluate these form factors by using dispersion relations and the heavy quark effective theory, 
it is convenient to redefine the form factors as follows \cite{CLN}: 
\begin{eqnarray}
 \label{V1} V_1(w) &=&h_+(w)-{1-r \over 1+r} h_-(w) \,,\\
 \label{S1} S_1(w) &=&h_+(w)- {1+r \over 1-r} {w-1 \over w+1} h_-(w) \,, \\
 \label{A1} A_1(w) &=& h_{A_1}(w) \,, \\
 \label{R1} R_1(w) &=& {h_V(w) \over h_{A_1}(w)} \,, \\
 \label{R2} R_2(w) &=& {h_{A_3}(w)+rh_{A_2}(w) \over h_{A_1}(w)} \,, \\
 \label{R3} R_3(w) &=& {h_{A_3}(w)-rh_{A_2}(w) \over h_{A_1}(w)} \,.
\end{eqnarray}

\subsection{Scalar and pseudo-scalar operators}
For the scalar and pseudo-scalar operators, the form factors are defined as 
\begin{eqnarray}
 \label{Eq:scalarD} \langle D(p_D)|\bar c b|\bar B(p_B)\rangle &=& \sqrt{m_Bm_D} (w+1) h_S(w) \ , \\
 \label{Eq:scalarDstar} \langle D^*(p_{D^*},\epsilon) | \bar c \gamma^5 b | \bar B (p_B) \rangle &=& -\sqrt{m_Bm_{D^*}}\, (\epsilon^* \cdot v ) h_P(w) ,
\end{eqnarray}
and then the hadronic amplitudes in Eqs.~(\ref{Eq:HAS1}) and (\ref{Eq:HAS2}) are represented as
\begin{equation}
 H_{S_1}^s =H_{S_2}^s =m_B\sqrt{r}(w+1)h_S(w) \,,
\end{equation}
in $\bar B \to D \tau \bar\nu$, and
\begin{eqnarray}
 H_{S_1}^\pm &=& H_{S_2}^\pm = 0\,,\\
 H_{S_1}^0 &=& -H_{S_2}^0 = -m_B \sqrt r \sqrt{w^2-1}\, h_P(w) \,,
\end{eqnarray}
in $\bar B \to D^* \tau \bar\nu$. 

Using the quark equations of motion we obtain 
\begin{eqnarray}
 \label{Eq:SEOM} i \partial_\mu \left ( \bar c \gamma^\mu b \right ) = \left(m_b - m_c \right) \bar c b \,, \\
 \label{Eq:PEOM} i \partial_\mu \left ( \bar c \gamma^\mu \gamma^5 b \right ) =- \left(m_b + m_c \right) \bar c \gamma^5 b,  
\end{eqnarray}
which lead to the following relations among form factors: 
\begin{equation}\begin{split}
 \label{Eq:SRelation} h_S(w) 
 &=h_+(w)- \frac{1+r}{1-r} \frac{w-1}{w+1} h_-(w) +O\left( \frac{\Lambda^2}{m_Q^2} \right) \\
 &=S_1(w) +O\left( \frac{\Lambda^2}{m_Q^2} \right) \ , 
\end{split}\end{equation}
\begin{equation}\begin{split}
 \label{Eq:PRelation} h_P(w) 
 &= \frac{1}{1+r} \left[ (w+1) h_{A_1}(w) +(rw-1) h_{A_2}(w) -(w-r) h_{A_3}(w) \right] +O\left( \frac{\Lambda}{m_Q} \right) \\
 &= \frac{A_1(w)}{1+r} \left[ w+1 -\frac{1-r^2}{2r} R_2(w) +\frac{\hat q^2(w)}{2r} R_3(w) \right] +O\left( \frac{\Lambda}{m_Q} \right), 
\end{split}\end{equation}
where we have used $m_{B(M)} = m_{b(c)} +\Lambda +O\left( \Lambda^2/m_{b(c)} \right)$. 
The absence of $1/m_Q$ corrections in Eq.~(\ref{Eq:SRelation}) is confirmed by the heavy quark expansion without resorting to the equations of motion. 
Thus we employ $h_S(w)= S_1(w) =\left[ 1+\Delta(w) \right] V_1(w)$, where $\Delta(w)$ involves the $1/m_Q$ corrections. 
On the other hand, there exist unknown $1/m_Q$ corrections in Eq.~(\ref{Eq:PRelation}). 
We ignore them, but use the axial vector form factors $A_1(w)$ and $R_{2,3}(w)$ including $1/m_Q$ corrections as described in Sec.~\ref{Sec:FF}. 
The $1/m_Q$ corrections in $R_2(w)$ are necessary to evaluate the rate of $\bar B\to D^*\ell\bar\nu$, 
because the experimental determination of the slope parameter $\rho_{A_1}^2$ assumes these corrections \cite{HFAG}. 
As for $R_3(w)$, since it does not contribute to $\bar B\to D^*\ell\bar\nu$ and no experimental data are available, 
we use the theoretical estimation by the heavy quark effective theory with $\pm100$\% error in $1/m_Q$ corrections.

\subsection{Tensor operator}
The form factors of the tensor operator are defined as 
\begin{eqnarray}
\label{Eq:tensorD1}
 \langle D(p_D)| \bar c \sigma^{\mu \nu} b |\bar B(p_B)\rangle 
 &=& -i \sqrt{m_Bm_D} h_T (w) (v^\mu v'^\nu - v^\nu v'^\mu )\ , \\
\label{Eq:tensorDstar} 
 \langle D^*(p_{D^*},\epsilon) | \bar c \sigma^{\mu \nu} b |\bar B (p_B) \rangle
 &=& -\sqrt{m_Bm_{D^*}}\, \varepsilon^{\mu\nu\rho\sigma} \Big [ h_{T_1}(w) \epsilon^*_\rho (v +v')_\sigma +h_{T_2}(w) \epsilon^*_\rho (v-v')_\sigma \notag \\
 &\,&+h_{T_3}(w) (\epsilon^* \cdot v) (v+v')_\rho (v-v')_\sigma \Big ] \,,
\end{eqnarray}
where $\sigma^{\mu\nu} =(i/2) \left[ \gamma^\mu, \gamma^\nu\right]$ and $\varepsilon^{0123}=-1$. 
The amplitudes corresponding to the operator $\bar c \sigma^{\mu \nu} \gamma_5 b$ are given by the following identity, 
\begin{equation}
 \sigma^{\mu \nu} \gamma_5 = -\frac{i}{2} \varepsilon^{\mu \nu \alpha \beta}\sigma_{\alpha \beta} \,.
\end{equation}
With the above form factors, the hadronic amplitudes in Eq.~(\ref{Eq:HAT}) are represented as 
\begin{eqnarray}
 H_{+-}^s(q^2) &=& H_{0s}^s(q^2)= - m_B\sqrt{r} \sqrt{w^2-1}\, h_T(w)\,, \\
 H_{\pm 0}^\pm(q^2)
 &=& \pm H_{\pm s}^\pm(q^2) \notag \\
 &=& \pm m_B \sqrt{r} \frac{1-rw\pm r\sqrt{w^2-1}}{ \sqrt {\hat q^2(w)} } \notag \\
 &&\hspace{1em} \times \left [ h_{T_1}(w)+h_{T_2}(w) +(w\pm\sqrt{w^2-1})(h_{T_1}(w)-h_{T_2}(w)) \right ] \,,\\
 H_{+-}^0(q^2)
 &=& H_{0s}^0(q^2) \notag \\
 &=& -m_B \sqrt{r} \left [ (w+1)h_{T_1}(w) +(w-1)h_{T_2}(w) +2(w^2-1)h_{T_3}(w) \right ] \,, \\ 
 H_{\lambda\lambda'}^{\lambda_M}(q^2) &=& -H_{\lambda'\lambda}^{\lambda_M}(q^2) \,,\\
 \text{others} &=& 0. 
\end{eqnarray}
As in the case of the scalar type operators, the equations of motion lead to
\begin{eqnarray}
\label{Eq:TEOM}
 \partial_\mu \left ( \bar c \sigma^{\mu\nu} b \right ) 
 &=& -\left(m_b + m_c \right) \bar c \gamma^\nu b -\left( i \partial^\nu \bar c \right) b +\bar c\left( i \partial^\nu b \right), 
\end{eqnarray}
and we obtain  
\begin{eqnarray}
 \label{Eq:TRelation}   h_T(w) &=&h_+(w)-\frac{1+r}{1-r} h_-(w) +O\left( \frac{\Lambda}{m_Q} \right), \\
 \label{Eq:T1Relation} h_{T_1}(w) &=& \frac{1}{ 2 \hat q^2(w) } \left[ (1-r)^2(w+1)h_{A_1}(w) - (1+r)^2(w-1) h_{V}(w) \right]+O\left( \frac{\Lambda}{m_Q} \right), \\
 \label{Eq:T2Relation} h_{T_2}(w) &=& \frac{(1-r^2)(w+1)}{ 2 \hat q^2(w) } \left[ h_{A_1}(w) - h_{V}(w) \right]+O\left( \frac{\Lambda}{m_Q} \right), \\
 \label{Eq:T3Relation} h_{T_3}(w) &=& -\frac{1}{2(1+r) \hat q^2(w)} \big[ 2r(w+1) h_{A_1}(w) -\hat q^2(w) \left(r h_{A_2}(w) -h_{A_3}(w) \right)\notag \\
 && \hspace{7em}-(1+r)^2 h_V(w) \big]+O\left( \frac{\Lambda}{m_Q} \right).
\end{eqnarray}
We find $h_T(w) =h_{T_1}(w) =\xi(w)$ and $h_{T_2}(w) =h_{T_3}(w)=0$ in the heavy quark limit.  
Similarly to the case of the scalar type operators, we ignore the unknown $1/m_Q$ corrections in the right-hand sides, and employ the vector and axial vector form factors with the $1/m_Q$ corrections.

\section{Decay rates}
\label{Ap:DR}
The differential decay rate $\bar B \to M\tau\bar\nu$ is represented as
\begin{equation}\label{Eq:DR}
d\Gamma^{\lambda_\tau}_{\lambda_M}
=\frac{1}{2m_B} \sum_l \left|\mathcal{M}^{\lambda_\tau,\lambda_M}_l (q^2,\cos\theta_\tau)\right|^2 d\Phi_3\,,
\end{equation}
where the three-body phase space $d\Phi_3$ is given by
\begin{equation}
 d\Phi_3=\frac{\sqrt{Q_+Q_-}}{256\pi^3 m_B^2}
         \left(1-\frac{m_\tau^2}{q^2}\right)dq^2d\cos\theta_\tau\,,
\end{equation}
and $Q_{\pm}=(m_B\pm m_M)^2-q^2$. 
Several decay rates used in the main text are obtained by integrating Eq.~(\ref{Eq:DR}) over $q^2$ and $\cos \theta_\tau$. 
For notational simplicity, we define the following quantities: 
\begin{equation}\label{Eq:tauhelicity}
 \Gamma^\pm (D) =\Gamma^{\pm1/2}_s,\quad \Gamma^\pm (D^*) = \sum_{\lambda_M=\pm1,0} \Gamma^{\pm1/2}_{\lambda_M}, 
\end{equation}
\begin{equation} \label{Eq:Dstar}
 \quad \Gamma (D^*_T) =\sum_{\lambda_M=\pm1} \sum_{\lambda_\tau} \Gamma^{\lambda_\tau}_{\lambda_M},
 \quad \Gamma (D^*_L) =\sum_{\lambda_\tau} \Gamma^{\lambda_\tau}_0.
\end{equation}

\end{document}